# Review of Superconducting Qubit Devices and Their Large-Scale Integration

**Hiu Yung Wong, Senior Member, IEEE**
San Jose State University, San Jose, CA 95112 USA

Corresponding author: Hiu Yung Wong (e-mail: hiuyung.wong@sjsu.edu).

This paper is the result of the work supported by the National Science Foundation under Grant No. 2125906, 2024 SJSU RSCA Seed Grant, 2024 SJSU RSCA Equipment Grant, and AMDT Endowment Funding from the College of Engineering, San Jose State University.

**ABSTRACT** The superconducting qubit quantum computer is one of the most promising quantum computing architectures for large-scale integration due to its maturity and close proximity to the well-established semiconductor manufacturing infrastructure. From an education perspective, it also bridges classical microwave electronics and quantum electrodynamics. In this paper, we will review the basics of quantum computers, superconductivity, and Josephson junctions. We then introduce important technologies and concepts related to DiVincenzo's criteria, which are the necessary conditions for the superconducting qubit technology to work as a useful quantum computer. Firstly, we will discuss various types of superconducting qubits formed with Josephson junctions, from which we will understand the trade-off across multiple design parameters, including their noise immunity. Secondly, we will discuss different schemes to achieve entanglement gate operations, which are a major bottleneck in achieving more efficient fault-tolerant quantum computing. Thirdly, we will review readout engineering, including the implementations of the Purcell filters and quantum-limited amplifiers. Finally, we will discuss the nature and review the studies of two-level system defects, which are currently the limiting factor of qubit coherence time. DiVincenzo's criteria are only the necessary conditions for a technology to be eligible for quantum computing. To have a useful quantum computer, large-scale integration is required. We will review proposals and developments for the large-scale integration of superconducting qubit devices. By comparing with the application of electronic design automation (EDA) in semiconductors, we will also review the use of EDA in superconducting qubit quantum computer design, which is necessary for its large-scale integration.

## I. INTRODUCTION

Quantum computers promise to solve some classical intractable problems effectively and efficiently by harnessing the principles of superposition, entanglement, and interference [1], [2], [3], [4], [5], [6]. This is possible because a quantum computer uses quantum bits (qubits) instead of classical bits. To build a successful quantum computer, the hardware needs to fulfill the five DiVincenzo's criteria [7], which require the hardware to provide and support the following: 1. A scalable physical system with well-characterized qubits; 2. The ability to initialize the state of the qubits to a simple fiducial state; 3. Long relevant decoherence times, much longer than the gate operation time; 4. A "universal" set of quantum gates; 5. A qubit-specific measurement capability. We will call them Criterion 1 to Criterion 5 in the following text.

DiVincenzo's criteria are just the necessary criteria. In reality, to build a useful quantum computer, a large-scale integration that supports seamless integration with quantum error correction is also required. This is because qubit states and qubit controls are analog in nature. Unlike the regenerative Complementary Metal-Oxide-Semiconductor (CMOS), they are sensitive to noise, gate infidelity, and error propagation [8]. Therefore, quantum error correction is required. It is expected that a fault-tolerant qubit (logical qubit) would be constructed from 100 to 1000 physical qubits [9], depending on the logical qubit error rate requirement and the physical qubit error rate. Moreover, to demonstrate quantum advantage, large problems are required, and it is expected that more than 1,000 logical qubits will be needed. Therefore, a useful quantum computer needs to have at least 1 million or even 100 million physical qubits [10], [11], [12], and large-scale integration is inevitable.

Currently, several technologies have demonstrated the integration of qubits on a small to medium scale. These include superconducting qubit [13], [14], silicon spin qubit [14], [15], photonic qubits [16], trapped ion qubit [17], [18], neutral atom qubit [19], and others. Among them, the superconducting qubit technology is one of the most promising ones. For example, IBM [20] and Google [21] have superconducting quantum computers with over 100 physical qubits. One-qubit and two-qubit gate error rates for most isolated systems have also reached ~$10^{-3}$. Combining with long qubit coherence time, they have reached the threshold of some error correction codes [22], [23]. A superconducting qubit quantum processing unit (QPU) is usually composed of superconducting microwave circuits and Josephson junctions [14]. The superconducting technology is promising because 1) it uses integration fabrication technology and can leverage the knowledge and infrastructure of CMOS and superconducting circuits (such as Rapid Single Flux Quantum, RSFQ [24]) processing techniques; 2) it has fast gate times in the order of tens of nanoseconds and has achieved a relatively long coherence times, more than hundred micro-seconds; 3) it uses Josephson junctions to create artificial atoms which is tunable and designable, enabling a large design space for engineering.

In this review article, we will discuss the history and development of various components of the superconducting qubit quantum computer and its integration strategies. Whenever possible, the review is written in a way to minimize the need for physics and mathematics knowledge, but is rigorous enough. In Section II, we will provide an overview of quantum computing and superconducting qubit quantum computer hardware (Criteria 1-5). In Sections III and IV, we will discuss the basics of superconducting electronics and the properties of the Josephson junction. In Section V, we will discuss various superconducting qubit realizations and their properties (Criteria 1, 3). In Section VI, we will discuss the coupling between qubits, which are essential to implement two-qubit entanglement gates (Criterion 4). In Section VII, we will discuss the readout strategies, Purcell filters, and quantum-limited amplifiers (Criterion 5). In Section VIII, we will discuss two-level system defects that are limiting the coherence time of the qubits (Criterion 3). Section IX will cover the applications of electronic design automation in superconducting qubit chip design. Finally, in Section X, we will discuss the methodologies for large-scale integrations, followed by conclusions in Section XI. A roadmap of this paper is given in Table I.

TABLE I
ROADMAP OF THIS PAPER AND THE PURPOSE OF EACH SECTION IN RELATIONSHIP TO DIVINCENZO'S CRITERIA AND LARGE-SCALE INTEGRATION

| | *II* | *III* | *IV* | *V* | *VI* | *VII* | *VIII* | *IX* | *X* |
|---|---|---|---|---|---|---|---|---|---|
| | Overview | Super-conducting Basics | Josephson Junction | SC Qubits | Two-qubit Gates | Readout | Two-level Defects | Electronic Design Automation | Large-scale Integration |
| DC 1 | ✓ | | ✓ | ✓ | | | | | |
| DC 2 | ✓ | | | | | | | | |
| DC 3 | ✓ | | | ✓ | | | ✓ | ✓ | |
| DC 4 | ✓ | | | | ✓ | | | | |
| DC 5 | ✓ | | | | | ✓ | | | |
| Integration | | ✓ | ✓ | | | | | ✓ | ✓ |

DC refers to DiVincenzo's Criteria. They are repeated here for convenience. DC 1. A scalable physical system with well-characterized qubits; DC 2. The ability to initialize the state of the qubits to a simple fiducial state; DC 3. Long relevant decoherence times, much longer than the gate operation time; DC 4. A "universal" set of quantum gates; DC 5. A qubit-specific measurement capability.

## II. QUANTUM COMPUTER OVERVIEW

A quantum computer utilizes superposition, entanglement, and interference to solve difficult problems [1]. A qubit can represent a general state, $|\Psi\rangle$, as a superposition of the basis states, $|0\rangle$ and $|1\rangle$. That is, $|\Psi\rangle = \alpha|0\rangle + \beta|1\rangle$, where $\alpha$ and $\beta$ are complex numbers. $|0\rangle$ and $|1\rangle$ are also called the classical states or the computational basis. We say that $|0\rangle$ and $|1\rangle$ span the computational space. The basis states are usually represented by two energy levels of a physical system (Criterion 1). However, they might not be limited to only two levels in a real physical system. In the operation process, higher levels ($|2\rangle$ and above) might be occupied. This is known as the leakage to non-computational space, resulting in errors. The energy difference between $|0\rangle$ and $|1\rangle$ in a physical implementation is the qubit energy, $E_{10}$. It can be translated to an equivalent frequency (qubit frequency, $\omega_q = 2\pi f_q$) using $E_{10} = hf_q = \hbar\omega_q$, where $h$ and $\hbar$, are the Planck constant and the reduced Planck constant, respectively. Superconducting qubits usually have a qubit frequency between 1 GHz and 10 GHz.

When a qubit is measured, its state will collapse to the measurement basis states (or classical states) (Criterion 5).

As a quantum object, qubits are very sensitive to the environment and can lose their coherence by coupling to a non-ideal environment (Criterion 3). Their robustness is commonly measured by two parameters, namely the energy relaxation time ($T_1$) and the dephasing time ($T_\phi$ or $T_2$) [14]. $T_1$ is a measure of how fast an excited state will decay to the ground state. $T_2$ is a measure of how fast a superposition state will lose its phase information. In experiments, $T_2^*$ is measured instead of $T_2$ and it is given by,

$$\frac{1}{T_2^*} = \frac{1}{2T_1} + \frac{1}{T_2}. \tag{1}$$

$T_1$ and $T_2$ ($T_\phi$ ) are associated with bit-flip and phase-flip error, respectively. To minimize bit-flip error, one needs to

design the qubit states so that the transition matrix between $|0\rangle$ and $|1\rangle$ due to the noise operator is small. And to minimize the phase-flip error, the qubit frequency should be independent of the noise [25].

In principle, any quantum circuit can be constructed or effectively approximated by a finite set of quantum gates, known as the universal sets of quantum gates [14], [26] (Criterion 4). Very often, such a set contains certain important one-qubit gates and one two-qubit entanglement gate. For example, $\{H, T, S, CNOT\}$ can infinitely approximate any quantum circuits. One-qubit gates are generally easy to implement. In a superconducting qubit, a one-qubit gate is usually achieved by applying an appropriate microwave pulse to the qubit capacitively. High gate fidelity (> 99.9%) has been achieved. Readers may refer to [14] for a more detailed treatment. In this paper, we will concentrate on the discussion of two-qubit gates only (Section VI).

Initialization of qubits is very important and non-trivial in quantum computers (Criterion 2). Many quantum algorithms require frequent reinitialization of qubits to the ground state, and reliable initialization is particularly important for quantum error correction. Qubits can be initialized in two ways, namely thermal reset and active reset. In thermal reset, the qubit will be idle for about ten times its $T_1$ so that excited qubits will decay to the ground state. Given that a superconducting qubit might have a $T_1$ of 100 μs and the gate time is only about 100 ns, thermal reset is considered to be too long as it corresponds to > 10,000 gate operations. In an active reset, the qubit will be measured first (readout). If it is in an excited state, a NOT gate will be applied to bring it to the ground state. This is much faster than thermal reset because the readout time is only in the order of hundreds of ns. In this article, we will not discuss initialization, although it should be noted that initialization and readout error, known as state-preparation-and-measurement (SPAM) error, is one of the dominating errors in quantum computers.

Fig. 1 shows the schematic of a typical superconducting qubit quantum computer. Microwave signals for reset, gate operation (qubit manipulation), and readout are generated using room-temperature electronics. The signals and thermal noise are attenuated when they reach the quantum chip so that the thermal noise is lower than the quantum noise at 10 mK. The signals interact with the qubit so that the qubit will evolve based on the Hamiltonian (the energy landscape in the system with the microwave signal) [14].

In a readout operation (Criterion 5), the signal will be transmitted through the feedline (co-planar waveguide, CPW). It is amplified by a quantum-limited amplifier (QLA) followed by further amplifications at higher temperature stages. Amplification is necessary because very few microwave photons are used in the readout process. QLA has the lowest noise figure (NF) and, thus, is put at the first stage of the amplification chain to minimize the overall noise figure due to Friis formula [27] (see also Eq. (20)). An amplifier amplifies signal and noise at the same time and inadvertently adds additional noise to its output. Therefore, the signal-to-noise ratio is reduced after each amplification. The amount of reduction is related to NF. So, it is important to minimize the NF of the amplification chain.

In this paper, we will only discuss a common readout mechanism for the most commonly used qubit, namely the transmon qubit. The qubit is coupled to a resonator. The resonant frequency ($\omega_r$), which is different from the qubit frequency ($\omega_q$), is shifted due to the coupling (dispersive shift). In this process, there is no energy exchange. The amount and direction of the shift depend on the state of the qubit. This shift is the cross-Kerr, $\chi$ [14], [28]. In Fig. 1, the signal will pass through when it is near the resonant frequency, resulting in a peak in the $S_{21}$ curve [29]. In simple words, $S_{21}$ is the ratio of port 2 (right) voltage to port 1 (left) voltage. It is a complex number as it contains phase information. When a signal is injected into port 1, $S_{21}$ can be regarded as the amount of signal successfully propagated to port 2. By measuring the phase of the transmitted signal, one can deduce the state of the qubit. This is a type of dispersive readout. Fig. 2 shows the real and imaginary parts of $S_{21}$ as a function of readout pulse frequency. In a design where the qubit frequency is less than the resonator frequency, an excited (ground) state will reduce (increase) the resonator frequency by about $|\chi|$.

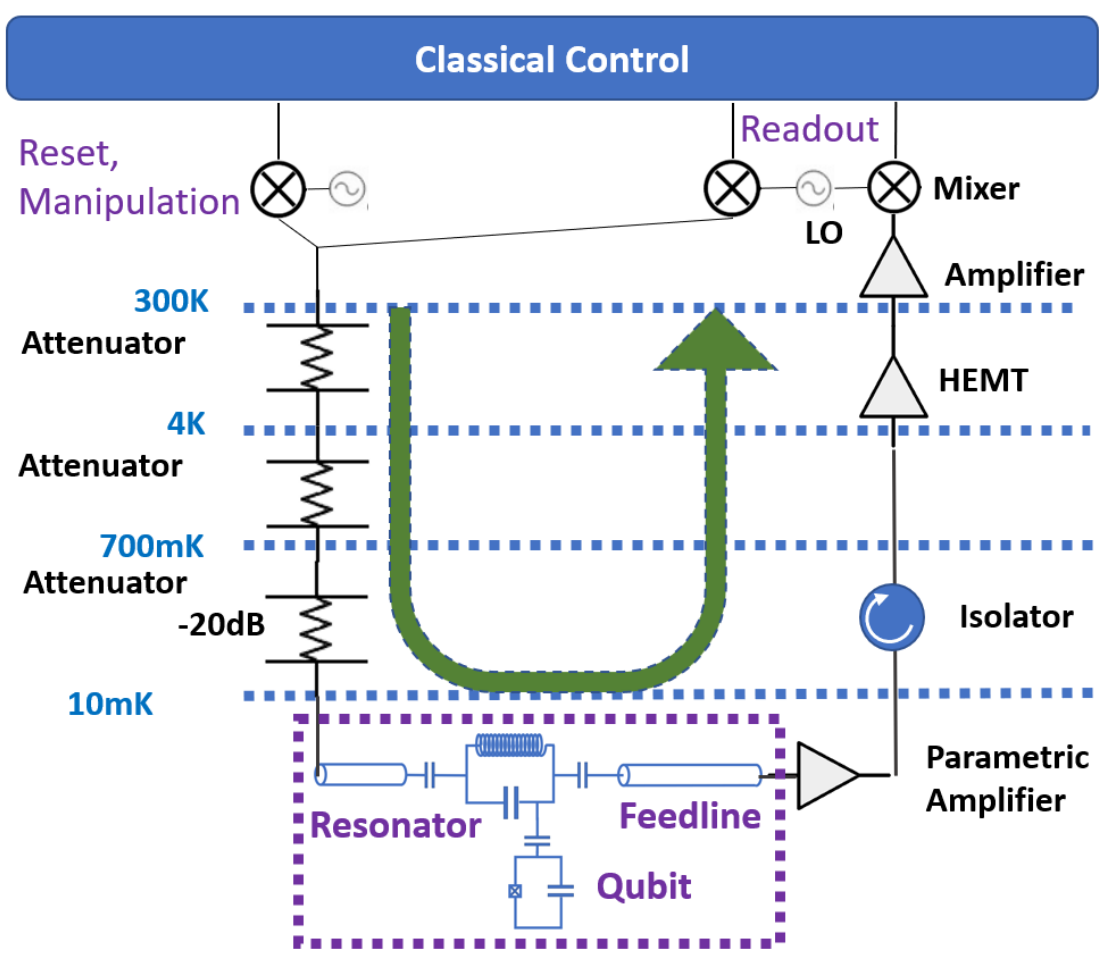

**FIGURE 1.** **A typical superconducting qubit quantum computer. Control signals are generated in classical electronics at room temperature. They are up-converted to microwave frequency (GHz) through a mixer and a local oscillator (LO) and sent to the qubit through various temperature stages with attenuations to reduce the thermal noise to below quantum noise, when it reaches the qubit chip. For readout signals, they will also go through an amplification chain with an isolator to avoid backflow of thermal noise. In this example, the amplification chain includes a parametric amplifier (a type of quantum-limited amplifier), a high electron mobility transistor (HEMT) amplifier at cryogenic temperature, and an amplifier at room temperature. Data are analyzed in classical electronics and computers. Only 1 qubit is shown. There may also be separate lines for flux control to tune the qubit frequency and qubit couplers. The components in the dashed box are usually integrated on a single chip.**

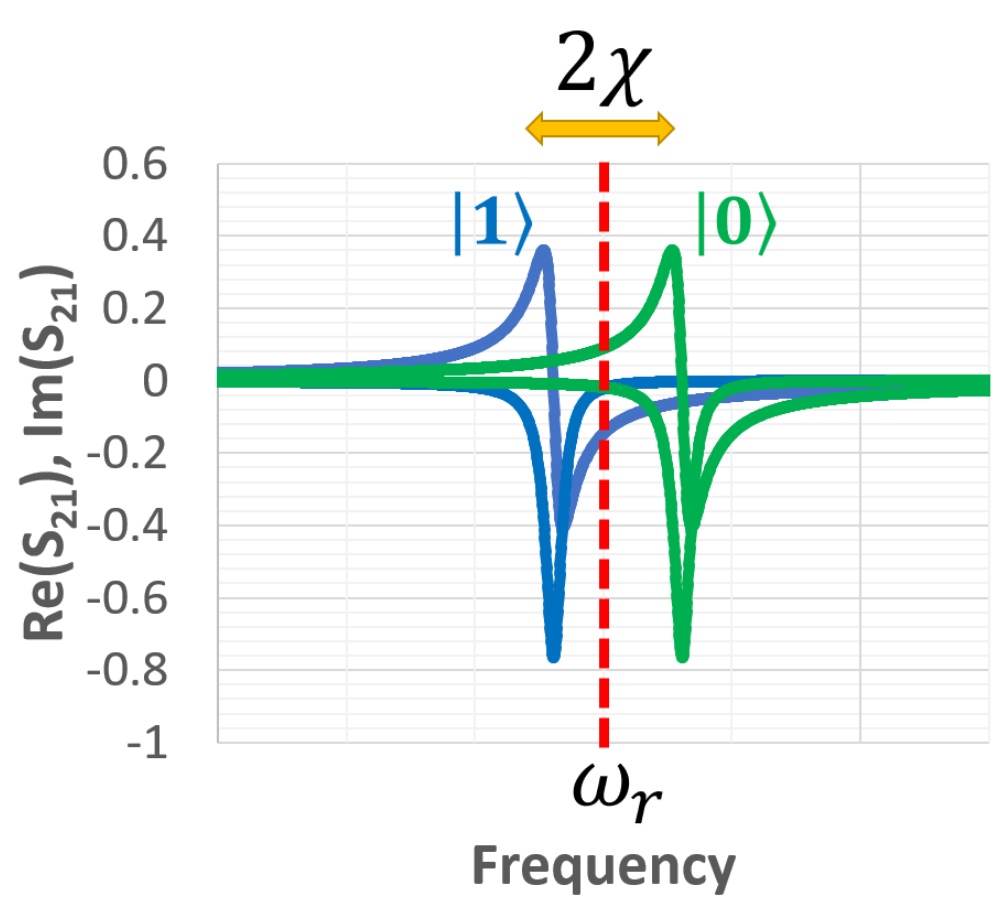

**FIGURE 2.** **Real and imaginary parts of $S_{21}$ when the qubit is at $|0\rangle$ and $|1\rangle$ state in a measurement. By measuring the phase at the bare resonant frequency, $\omega_r$, the qubit state can be deduced based on the real and imaginary values (i.e., the phase) of $S_{21}$.**

## III. SUPERCONDUCTING BASICS

For some materials, it becomes superconducting when its temperature, $T$, is below its critical temperature, $T_c$. At its superconducting state, its resistivity becomes zero. More importantly, unlike a perfect conductor, which has zero resistivity, it will also expel any magnetic field that has been trapped in the material when it transits from the normal state to the superconducting state. This is called the Meissner effect [14], [30], [31]. But it cannot expel an arbitrarily large magnetic field. Its superconducting state will be destroyed when an external magnetic field, $H$, is above a certain value, known as the critical magnetic field, $H_C$. $T_c$ reduces when the external magnetic field increases. $H_0$ is the external magnetic field above which $T_c$ becomes 0 K (i.e., $H_0 = H_C(T = 0K)$) (Table II). It should be noted that these critical values are defined when there is no current flowing through the material. At a given $T$ and $B$, the superconducting state will be destroyed when the current or current density is above a certain value known as the critical current, $I_c$, or the critical current density, $J_c$. Aluminum (Al) [32], niobium (Nb) [33], niobium nitride (NbN) [34], and tantalum (Ta) [35] are some commonly used superconducting materials in quantum computers. Their critical temperatures are 1.2 K [36], [37], >9 K [38], 16 K [39], and >4.2 K [38], respectively (Table II). It should be noted that $T_c$ depends on the thickness, quality, and fabrication process of the films. It may increase [36] or decrease [39] as the film thickness decreases. For example, $T_c$ of epitaxial NbN thin films on (100) MgO changes from 16 K to 11.5 K when the film thickness reduces from 100 nm to 3.3 nm [39].

Due to the Meissner effect, superconductors expel an external magnetic field by forming screening supercurrents near the surface. There is a finite magnetic field penetration depth, $\lambda$, since the screening current cannot be infinite (Table II). Moreover, thinner films are found to have a larger $\lambda$. For example, for Al, $\lambda$ increases from 58 nm to 163 nm when the thickness, $d$, reduces from 207 nm to 28 nm [40].

TABLE II
PROPERTIES OF COMMON SUPERCONDUCTING MATERIALS

| | $T_c$ (K) | $H_0$ (T)[#] | $\lambda$ (nm) | $\Delta$(T=0) (meV)[*] |
|---|---|---|---|---|
| Al | 1.2 [36] | 0.01[37] | >58[40] | 0.18 |
| Nb | 9 [38] | >2.8[41] | >43 [42] | 1.37 |
| NbN | 16 [39] | 35[43] | >300[44] | 2.43 |
| Ta | >4.2[38] | 0.083[45] | >100[46], [47] | 0.64 |

The properties are film-thickness- and film-quality-dependent. Some of them are thin-film values. This table represents examples of what has been achieved. Readers should carefully check the experimental conditions. [#]Nb and NbN are type II superconductors, and their upper critical fields are quoted [30]. [*] Calculated using Eq. (2) with the $T_C$ values in the first column.

At its superconducting state, electrons with opposite spin and momentum pair up as Cooper pairs, which have zero spin and act as bosons [30], [31]. Their spatial extension is called the coherence length, $\xi_0$. For example, $\xi_0$ of bulk Al and Nb are about 1500 nm [48] and 40 nm [49], respectively. The Cooper pairs are more energetically favorable by a gap energy, $\Delta$(T), than being broken (quasi-particle excitation) (Table II). The energy, $E_g(T)$, required to excite a Cooper pair to two quasi-particles is thus, $2\Delta$(T). Note that they are temperature-dependent, and they are related to $T_c$ by [30],

$$E_g(T=0) = 2\Delta(T=0) = 3.528kT_c, \quad (2)$$

where $k$ is the Boltzmann constant. As bosons, Cooper pairs can occupy the same ground state like photons in a laser. This is a macroscopic quantum object (superconducting condensate) and can be described by an order parameter in the theory of phase transition [30]. The order parameter, $\psi$, is just the macroscopic wavefunction of the superconducting condensate, and can be expressed as

$$\psi = |\psi|e^{i\theta}, \quad (3)$$

where θ is the phase. It should be noted that, while a global phase in a microscopic wavefunction is not measurable and the relative phase can only be measured via interference, the phase in a superconducting condensate is meaningful, and its difference is directly measurable (see Section IV). The square of the magnitude of the order parameter, $|\psi|^2$, is the Cooper pair density. Therefore, the phase describes the collective condensate of ~$10^{23}$ Cooper pairs per $cm^3$.

Some superconductors also have large kinetic inductance [31]. The source of kinetic inductance is due to the inertia of the moving Cooper pairs. This is different from geometric (magnetic) inductance and can be very large for some superconductors. Kinetic inductance is nonlinear as it depends on the strength of the supercurrent. When $\lambda$ is larger than the film thickness $d$, it is known that the kinetic inductance will dominate over the geometric inductance. Therefore, a thinner film has a larger kinetic inductance [50]. The specific kinetic inductance of a thin film, $L_K$, is given by

$$L_K = \mu_0\lambda^2/d, \quad (4)$$

where $\mu_0$ is the vacuum permeability. Therefore, Nb has a small kinetic inductance due to its small $\lambda$. On the other hand,

NbN has a large kinetic inductance, but its $\lambda$ also depends on its stoichiometry [50].

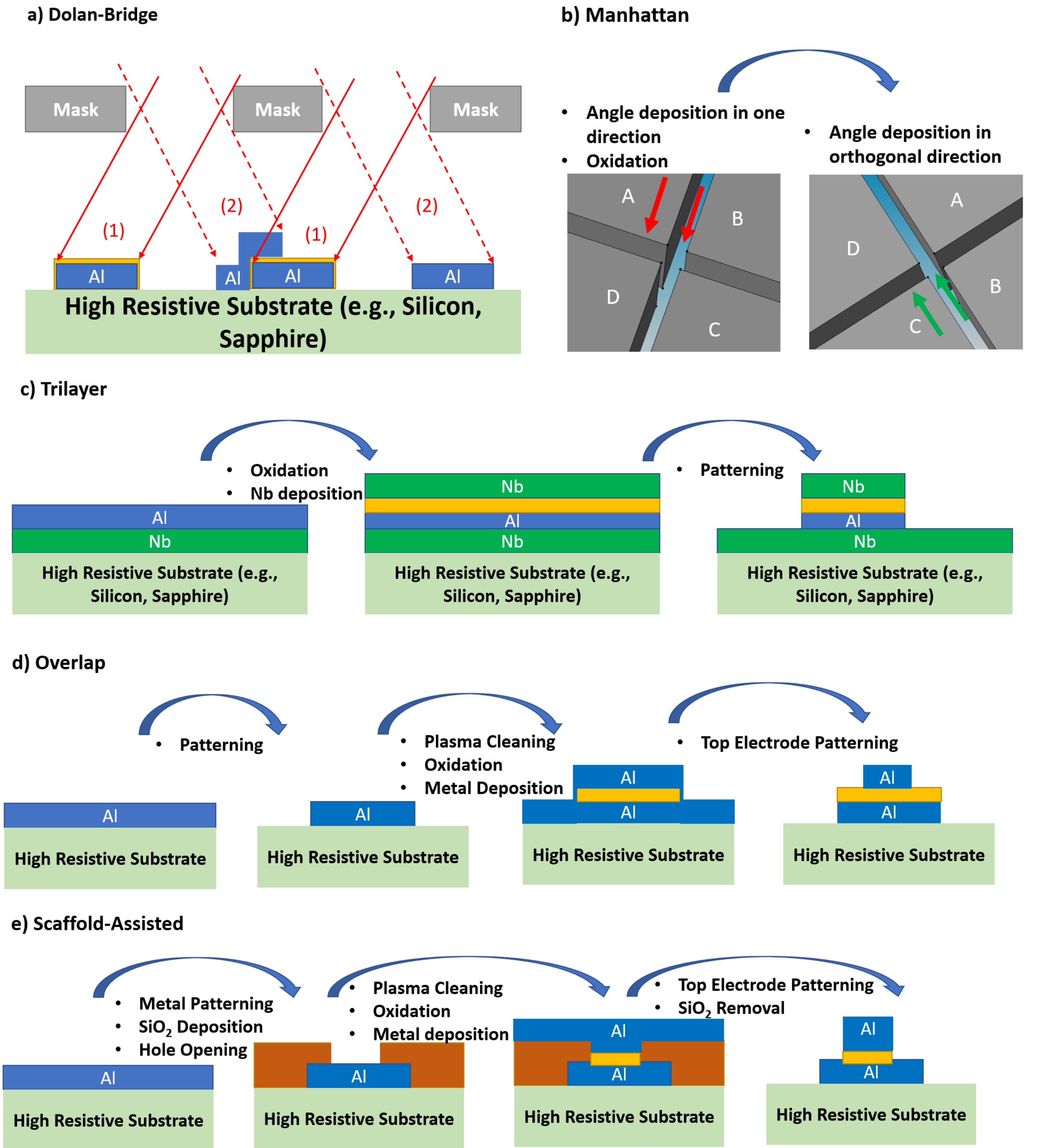

**FIGURE 3.** Several Josephson junction fabrication processes. The yellow regions are tunneling/ junction barriers, usually $AlO_x$. Brown is $SiO_2$. Here, Al and Nb are the superconductors. Major processing steps are highlighted. Processing sequences are either indicated by step numbers (Dolan bridge) or arrows (other methods). a) Dolan bridge: suspended photoresist is formed. Angle deposition is performed in step (1) followed by oxidation, and then another angle deposition in step (2). b) Manhattan: thick photoresist is patterned like the tall buildings in Manhattan. Angle deposition along one opening (street) is performed, followed by oxidation, and then another angle deposition along the orthogonal opening (cross street). Due to shadowing, a Josephosn junction is formed only at the crossing. c) Trilayer: uniform deposition of Al/Nb and oxidation of Al is possible on a blanket wafer. The junction is then patterned by etching. d) Overlap: it is a CMOS-compatible process. e) Scaffold-assisted: similar to the overlap process, it is CMOS compatible. A sacrificial oxide is used to avoid damaging the substrate during plasma cleaning. The sacrificial oxide is then removed using vapor HF.

## IV. JOSEPHSON JUNCTION AND ITS FABRICATION

We will first discuss the Josephson junction (JJ) (Fig. 3). A Josephson junction is just a superconductor-insulator-superconductor stack. Each superconducting electrode has its own order parameter, $\psi_i$, and the corresponding phase, $\theta_i$. The insulator is thin enough to allow tunneling of the Cooper pairs and thus serves as a weak link between the two macroscopic wavefunctions. Therefore, the insulator is called the tunneling barrier or junction barrier. This results in two Josephson equations that relate the junction current density, $J$, and junction voltage, $V$, to the junction phase, $\varphi = \theta_1 - \theta_2$, as

$$J = J_c \sin\varphi, \tag{5}$$

$$\frac{\partial\varphi}{\partial t} = \frac{2e}{\hbar}V, \tag{6}$$

where e and $t$ are the elementary charge and time, respectively. Therefore, the junction current depends on the phase difference between the superconducting condensates in the two electrodes. That is why in Section III, we say that the phase difference is measurable. It should be noted that a Josephson junction can be modeled as a bare junction that obeys Eqs. (5) and (6) and a parallel junction capacitor, $C_J$, since it has a conductor/insulator/conductor structure (top right in Fig. 4).

The Josephson junction equations do not depend explicitly on the tunneling barrier thickness and electrode material, which are encapsulated in $J_C$. When $J < J_C$, the Josephson junction will conduct a constant current with $V = 0$. It should be noted that, therefore, the critical current density of a Josephson junction is the current density below which the Josephson effect can be maintained, which is usually less than the critical current that will stop the superconductivity of the metal. Based on the Ambegaokar–Baratoff relation, it is known that $J_c$ is related to the room temperature tunneling resistance, $R_n$, through [51],

$$J_c = \frac{\pi\Delta}{2R_n eA}, \tag{7}$$

where $A$ is the junction area. Therefore, $J_c$ is very sensitive to the thickness of the tunneling barrier.

A Josephson junction acts as a nonlinear inductor and also stores energy like a regular geometric inductor. Its inductance, $L$, and energy, $E$, are given by [14],

$$L = \frac{\Phi_0}{2\pi I_c \cos\varphi} = \frac{L_J}{\cos\varphi}, \tag{8}$$

$$E = -\frac{I_c\Phi_0\cos\varphi}{2\pi} = -E_J\cos\varphi, \tag{9}$$

where $\Phi_0$, $L_J$, and $E_J$ are the magnetic flux quantum, Josephson inductance, and Josephson energy, respectively. The magnetic flux quantum can also be expressed as $\Phi_0 = h/(2e)$. Therefore, the properties of a Josephson junction also depend on $A$, because $I_c = AJ_c$. It should be noted that Josephson inductance is predominantly kinetic inductance due to the current flowing through the weak link. Based on Eqs. (8) and (9), a large $I_c$, results in a small $L_J$ and a large $E_J$. A large $I_c$ also means a large area (for the same $J_c$ or technology) and thus a large capacitance, which is translated to a small capacitive energy. These properties will be used to design some qubits in Section V. For example, a larger area can be used to increase the inductive energy and reduce the capacitive energy.

A commonly used Josephson junction in quantum computing has an Al/$AlO_x$/Al stack [52], [53], where $AlO_x$ refers to sub-stoichiometric aluminum oxide.

Josephson junctions are commonly fabricated using the Dolan Bridge [14], [54], or the Manhattan processes [14], [55] (Fig. 3a and b). In the Dolan bridge method, bilayer resist is needed in order to form a suspended bridge [56]. Both of them rely on a double-angle shadow evaporation with in-situ oxidation without breaking the vacuum, while the photoresist is still there. In the processes, an angle deposition of the metal for one electrode is performed, followed by metal oxidation (yellow) in the presence of photoresist and a final angle deposition of the second electrode. Photoresist liftoff is also needed to remove unwanted metal. This is uncommon in semiconductor processes and is suspected of causing defects and thus reducing qubit coherence time when they are used to construct a superconducting qubit (see Section VIII). The uniformity and reproducibility are also not as good as other standard CMOS processing steps [57]. It cannot leverage the state-of-the-art CMOS process for large-scale integration. On the other hand, such processes require no etching and thus can avoid damage-induced defects.

To improve uniformity, a trilayer stack using Nb/Al-$AlO_x$/Nb has been proposed to form the Josephson junction (Fig. 3c) [58], [59]. Nb (100 nm)/ thin Al deposition, followed by oxidation and Nb (100 nm) deposition, is performed before seeing the photoresist. Advanced lithography tools can be used to pattern a highly uniform trilayer stack. This provides better control of the tunneling barrier thickness and junction area.

Another approach to use a standard CMOS process is the overlap junction (Fig. 3d) [52] (There are other variations, e.g., in [60]). In an overlap process, after the first metal is deposited and patterned, it is cleaned by Ar plasma. After that, it is oxidized, followed by the deposition of the second metal without breaking the vacuum. Despite using plasma cleaning and etching processes, it shows the possibility of achieving a high coherence time.

The overlap junction method has a potential issue that during the in-situ Ar plasma cleaning of the first metal layer, defects may form on the exposed substrate not covered by the metal. Such defects can be detrimental to qubit coherence time (to be discussed in Section VIII). Therefore, in [61], it was proposed to use $SiO_2$ as a scaffold (Fig. 3e). After the deposition and patterning of the first metal, $SiO_2$ is deposited and patterned to open the junction area. The in-situ Ar plasma cleaning is performed, during which no substrate is exposed to the plasma. It is followed by in-situ oxidation and second metal deposition. At the end of the process, the top electrode is patterned, and $SiO_2$ is removed using vapor HF. A promising coherence time of 57 µs has been achieved.

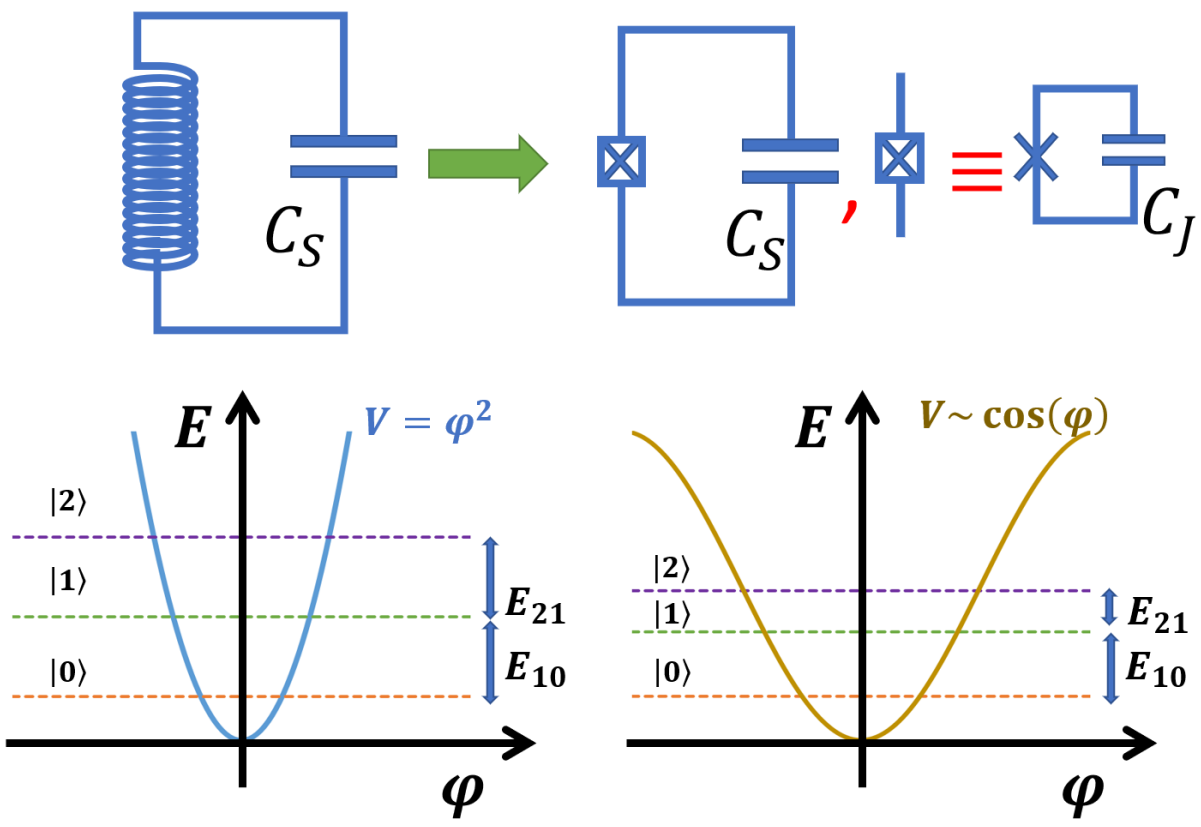

FIGURE 4. Left: An LC tank circuit (top) and its energy in the phase space, which has zero anharmonicity (bottom). Right: A Josephson junction replaced the linear inductor in an LC tank (top). A Josephson junction has a bare junction that obeys the Josephson equations and a parallel junction capacitor (top right). Its symbol is a cross in a square, which includes the bare junction and the junction capacitor. Its energy in the phase space is non-parabolic (shifted cosine function), which has substantial anharmonicity (bottom). Note that the potential is periodic in $\varphi$.

To enable large-scale integration, fabrication of Josephson junctions by leveraging state-of-the-art CMOS processes is desired so that better junction area and thickness uniformity, and reproducibility can be achieved. 2D materials have also been considered to be used for forming Josephson junctions [62] to avoid the thickness inhomogeneity of $AlO_x$. They provide precise control of the oxide thickness and avoid dangling bonds at the superconductor/insulator interfaces [63].

There are other efforts to modify the fabrication process in order to achieve a higher quality Josephson junction. For example, a rhenium (Re) /crystal-$Al_2O_3$/Al Josephson junction is grown by annealing the deposited amorphous $AlO_x$ to crystalline $Al_2O_3$, in order to reduce the defects in the $AlO_x$ barrier [64]. Similarly, epitaxial growth of NbN/AlN/NbN was also experimented with to avoid defects in the amorphous tunneling barrier [65].

Josephson junction variations are a critical barrier to large-scale integration. To alleviate this problem, laser annealing has been proposed to adjust the qubit frequency after fabrication in [66] to achieve precise settings of qubit frequencies. Junction area fluctuation can also be minimized by optimizing the stepping field size, step size, and evaporation angle. It is shown that a smaller field size and angle result in less line edge roughness (LER), and thus less junction area variations [56].

## V. SUPERCONDUCTING QUBITS

In this section, we will discuss common superconducting qubits. All of them have Josephson junctions as components to provide non-linearity. A qubit is formed by combining Josephson junctions with external capacitors and inductors (e.g., Fig. 4). The junction capacitor also plays a critical role when there is no external capacitor, such as in the charge and flux qubit cases. We will start with the discussion of the charge qubit and the transmon qubit. After that, we will discuss the development of the phase qubit, flux qubit, and fluxonium qubit. Finally, we will discuss other types of qubits, such as the gatemon qubit and the cat qubit. We will also discuss how to achieve a flux-tunable qubit. In [67], a comparison of various types of superconducting qubits in the early days was given.

Qubit design is a trade-off between the parameters of the Josephson junction (mainly $L_J$ or $E_J$), the capacitor (thus $C$ and the charging energy, $E_c$), and the inductor (thus, $L$ or $E_L$). The definitions of $E_c$ and $E_L$ are shown in Fig. 5. Fig. 5 also shows the relative design parameter values of different kinds of superconducting qubits based on the data in [67] and [25]. Superconducting qubits are treated as artificial atoms. This is because quantized energy levels can be engineered, and they can be coupled to a resonator/cavity, like how a natural atom coupled to an optical cavity (see also Sections VI and VII).

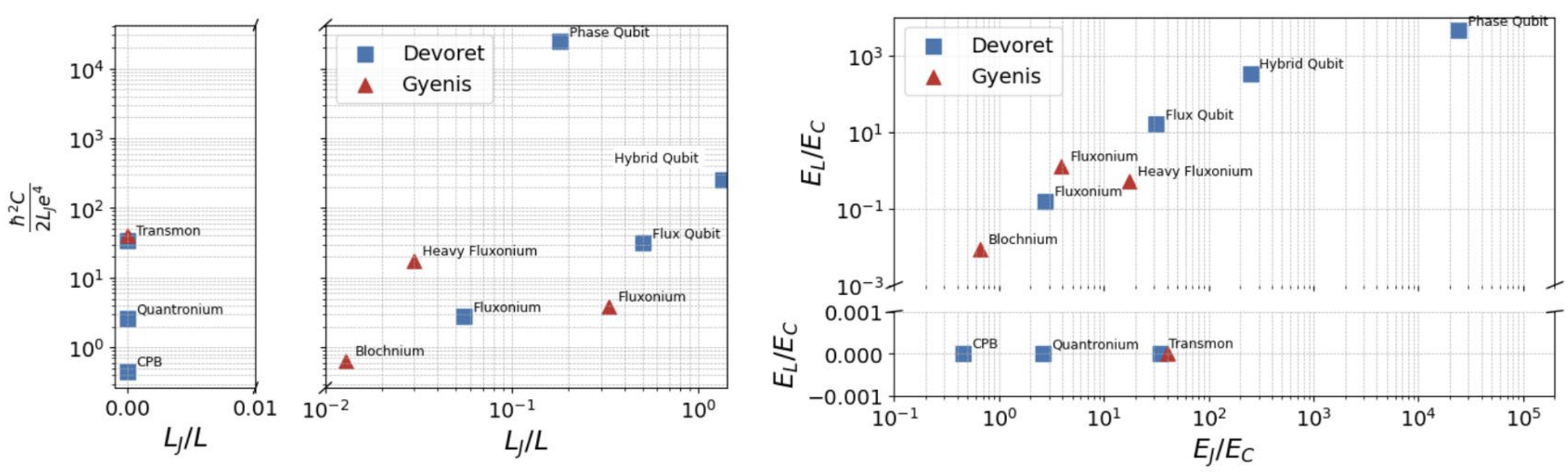

FIGURE 5. Plots of different types of qubits as a function of various quantities. Data are digitized from Devoret et al. [65] and Gyenis et al. [25]. The left figure uses the ratios given in Devoret et al. [65], and the right figure uses the ratios given in Gyenis et al. [25]. The following equations were used for conversion: $E_J = \frac{(\Phi_0/2\pi)^2}{L_J}$ (based on Eqs. (8) and (9)), $E_L = \frac{(\Phi_0/2\pi)^2}{L}$, and $E_C = \frac{e^2}{2C}$. Therefore, $\frac{E_J}{E_C} = \frac{\hbar^2 C}{2L_J e^4}$ and $\frac{E_L}{E_C} = \frac{\hbar^2 C}{2L_J e^4}\frac{L_J}{L}$.

Therefore, Fig. 5 can be referred to as a "Mendeleev-like but continuous table of artificial atom types" [67]. We will refer to this table when we discuss various types of qubits, although not all qubits in the figure are covered.

### A. L-C TANK, CHARGE QUBIT, AND TRANSMON

Classically, an L-C tank has a resonant frequency, $\omega_0$, and it can store microwave energy. The Hamiltonian (which is the total energy in this case) of an L-C tank has the same form as a mechanical simple harmonic oscillator (SHO) [14], [68]. Therefore, its energy levels can be quantized like a mechanical SHO (Fig. 4). At cryogenic temperature (e.g., 10 - 20 mK at which a superconducting qubit usually operates, Fig. 1), its energy levels are well-separated and well-defined (Criterion 1). Level $N$ corresponds to having $N$ quanta of energy due to $N$ microwave photons and has an energy of $E_N = \hbar\omega_0(N + 1/2)$. Therefore, $N$ is a non-negative number. These states are also called the Fock state, $|N\rangle$. The energy levels are evenly separated and, thus, there is a lack of anharmonicity. Anharmonicity, $\alpha$, is defined as the difference between the $|2\rangle/|1\rangle$ energy spacing, $E_{21} = E_2 - E_1$, and $|1\rangle/|0\rangle$ energy spacing, $E_{10} = E_1 - E_0$. That is,

$$\alpha = E_{21} - E_{10}. \tag{10}$$

As mentioned in Section II, a quantum gate is the application of a microwave pulse. In an LC-tank, while we may use level 0 as state $|0\rangle$ and level 1 as state $|1\rangle$ to form a qubit, $|1\rangle$ can be driven to $|2\rangle$ easily, when it is supposed to be driven to $|0\rangle$, due to the lack of anharmonicity ($\alpha = 0$). Therefore, an LC-tank is not a good qubit because the computational space cannot be restricted to the bottom two levels (violation of Criterion 1). Moreover, it is desirable to have a large anharmonicity. This is because any finite pulse has other undesired frequency components, which can drive the qubit to non-computational space. A shorter pulse has more such components. Therefore, to use a shorter pulse (thus a faster gate) without substantial leakage, a larger anharmonicity is needed.

As mentioned in Section IV, a Josephson junction acts as a non-linear inductor without introducing dissipation to the superconducting circuit. By replacing the linear inductor in an L-C tank with a Josephson junction, one can achieve the desired anharmonicity (i.e., energy levels are not evenly spaced) (Fig. 4). This is due to the cosine function in Eq. (9). The capacitor can be a standalone capacitor or the junction capacitor of the Josephson junction.

Here, we consider a more general case in which the system is also capacitively coupled to a voltage source, which is also known as a Cooper pair box (CPB) [69], [70] (Fig. 6). The Hamiltonian of the system is found to be [14],

$$H = 4E_c(n - n_g)^2 - E_J \cos\varphi, \tag{11}$$

where $E_c = e^2/(2C_\Sigma)$, $n$, $n_g$, are the electron (not Cooper pair, which has a charge of $-2e$ instead of $-e$ ) charging energy, number of excess Cooper pairs on the island (between JJ and $C_g$), and the number of Cooper pairs induced by the coupling capacitance, respectively. $C_\Sigma$ is the total capacitance, including the Josephson junction self-capacitance. $n$ refers to the number of excess Cooper pairs and thus can be any integer (unlike $N$). To understand the behavior of this system, we need to promote $H$, $n$, and $\varphi$ to operators and then diagonalize the Hamiltonian operator to represent the system in the eigen-basis of the Hamiltonian [14]. Its first and second lowest energy levels will then be used to represent state $|0\rangle$ and state $|1\rangle$ of a qubit.

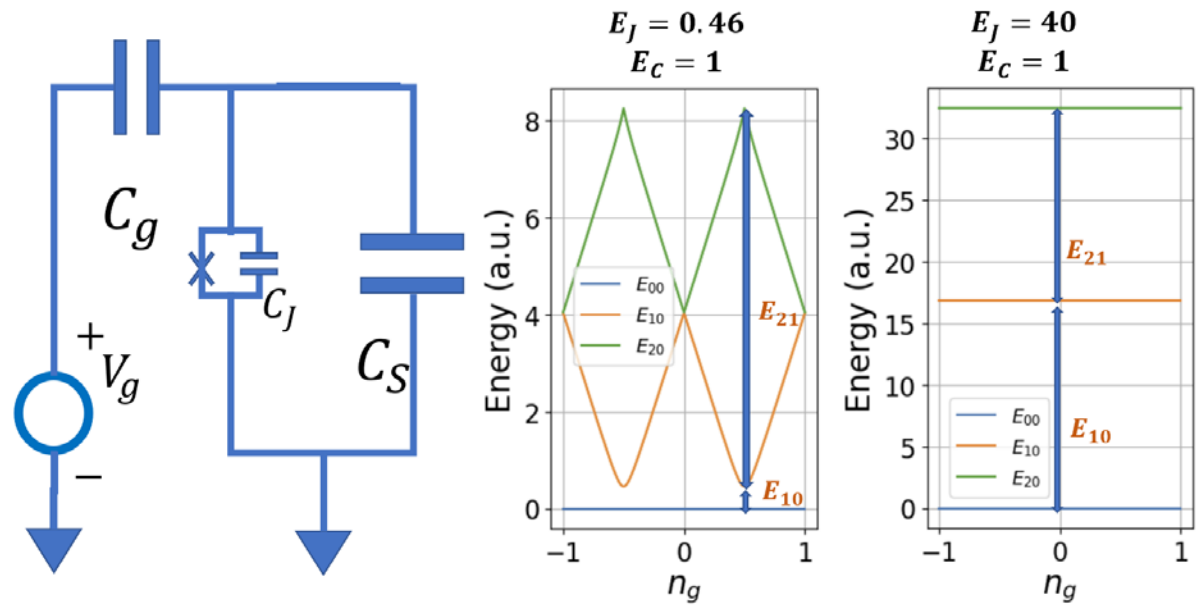

**FIGURE 6.** **Left: A Cooper Pair Box (CPB) coupled to an external voltage source $V_g$. It is the same as the circuit in Fig. 4 with the addition of an external voltage source. Therefore, the energy landscape in the phase space is the same as that in Fig. 4. Middle: The energy level spacing as a function of offset charge $n_g$ for the CPB in Fig. 5. Right: The energy level spacing as a function of offset charge $n_g$ for the transmon in Fig. 5. It is clear that while the charge sensitivity is improved in a transmon, the anharmonicity is also reduced.**

Compared to a mechanical SHO, $n$ and $\varphi$ can be regarded as a scaled generalized momentum and position, respectively. Therefore, they obey the Heisenberg uncertainty principle. When $n$ is well defined, $\varphi$ has a high uncertainty, and vice versa. The first and the second terms in Eq. (11) are the corresponding kinetic and potential energy, respectively. Due to $\cos\varphi$, the potential well in the $\varphi$ space is periodic. The $\cos\varphi$ term provides anharmonicity. To have substantial anharmonicity, the quantized levels should be high so that they are not near the bottom of the well, where it is approximately parabolic. That means $E_c$ needs to be large compared to $E_J$ (Fig. 5). We can also understand it from another angle. When $E_c$ is large, $n$ is well-defined (as it penalizes the change of $n$). Therefore, it has more uncertainty in $\varphi$ and thus can sample more of the cosine nonlinearity.

One of the special cases is when $n_g = 0.5$ (the sweet spot) and $E_c \gg E_J$ [14], [69], which is usually the case if no standalone parallel capacitor is added ($C_S = 0$) (Fig. 6 left and middle). The system has a very large anharmonicity and can be used as a qubit. But it should be noted that, in this case, $|0\rangle = (|n=0\rangle + |n=1\rangle)/\sqrt{2}$ and $|1\rangle = (|n=0\rangle - |n=1\rangle)/\sqrt{2}$ and therefore, the eigenstates of a charge qubit are not just $n$ (excess Cooper pairs on the island) but a superposition of them. However, the charge number still has a relatively small spread (well-defined), and thus it is called a charge qubit. It resembles the bonding and anti-bonding states in a double-well quantum dot. Moreover, its eigenstates are also not those of an L-C tank, which are just the number of photons (Fock states, $|N\rangle$). Since the Hamiltonian depends on $n_g$, which can be induced easily by charge noise through capacitive coupling, a charge qubit is very sensitive to charge noise even at the sweet spot (Fig. 6). Therefore, a charge qubit usually has low $T_1$ ($< 5\mu s$) (e.g., [71]). While $T_1 \sim 200\mu s$ has been achieved with special circuit design and engineering, its $T_2^*$ is $< 0.5\mu s$ [53] because $E_{10} = \hbar\omega_q$ changes strongly with $n_g$.

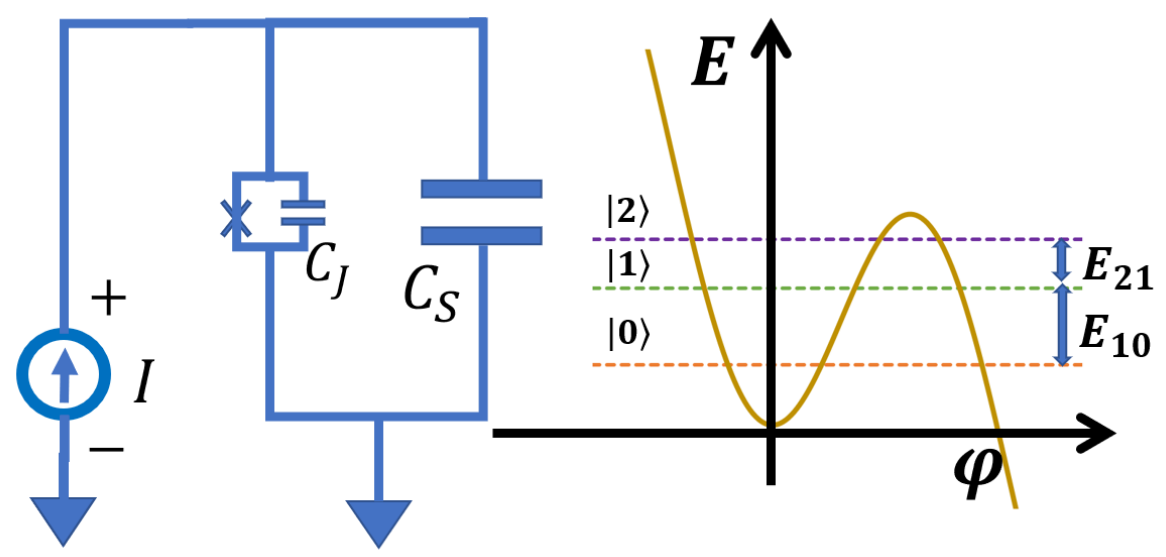

FIGURE 7. A charge qubit is biased with a constant current. As a result, the periodic cosine potential in Fig. 4 is tilted in the phase space. This allows a large anharmonicity even with a large capacitance. |2⟩ can also be used for readout, as it can tunnel through the barrier in the phase space easily. Readout is performed by applying $E_{21}$ to the qubit. If it is at state |1⟩, it will be excited to |2⟩ and the tunneling can be measured (as a potential drop across the junction due to Eq. (6)). There will be no potential drop if it is at the state |0⟩.

To minimize charge sensitivity, one may operate it in the regime of $E_c \ll E_J$ (Fig. 6 right). The second term in Eq. (11) will then dominate, and the effect of charge noise due to $n_g$ will reduce. This can be achieved by adding a large shunt capacitor, $C_S$, to increase $C_\Sigma$, to reduce the charging energy, or in other words, to increase the Josephson energy to capacitive energy ratio, $E_J/E_C$. This device is called a transmon [28]. However, at the same time, the anharmonicity is also reduced. Fortunately, its charge sensitivity is suppressed exponentially at a rate of $e^{-\sqrt{E_J/E_C}}$ and the anharmonicity only reduces as $\sqrt{E_J/E_C}$ [72]. The transmon qubit is the most commonly used superconducting qubit for large-scale integrations [9], [13], [21], [73], [74]. It has achieved high $T_1$ (hundreds of microseconds or even $> 1ms$ with appropriate engineering) [35] [75] and $T_2^*$ times [25], [75], [76]. A transmon qubit is a charge-mode qubit, and it has a long $T_2$ as its frequency is insensitive to charge noise. Its $T_1$ is currently limited by two-level-system defects (see Section VIII) and other losses.

Since a transmon qubit has less anharmonicity (Fig. 6), it can be approximated as an L-C tank oscillator with an additional non-linear term ($-E_J\varphi^4/24$) [14]. In other words, the Taylor expansion of $cos\,\varphi$ in Eq. (11) is kept up to the fourth-order term, and the fourth-order term is used as a perturbation. Therefore, its lowest two levels are similar to the lowest two levels of a quantized L-C tank mathematically and physically. They are the states with $N = 0$ and $N = 1$ photon (Fock states), with $E_N = \hbar\omega_0(N + 1/2)$, respectively. They are used to represent state $|0\rangle$ and state $|1\rangle$ of a qubit. Of course, they are also a linear superposition of excess Cooper pairs $|n\rangle$. The qubit frequency $f_q = \omega_q/2\pi$ is usually 3 GHz to 8 GHz and can be found using,

$$\hbar\omega_q = \sqrt{8E_cE_J} - E_c. \qquad (12)$$

Its anharmonicity is given by $E_c$ which is in the order of 100 MHz.

### B. Phase Qubit, Flux Qubit, and Fluxonium

A phase qubit is a Josephson junction biased with a constant current source, $I$, (Fig. 7) [77]. This is also the system used to demonstrate macroscopic tunneling and energy quantization in an electrical circuit, leading to the 2025 Nobel Physics Prize [78]. The Hamiltonian of the circuit is given by,

$$H = 4E_cn^2 - E_J\cos\varphi - \frac{I\Phi_0\varphi}{2\pi}. \qquad (13)$$

This is the same as Eq. (11) except that the offset charge is not included, and has an additional term due to the current source. The last term tilts the potential well, so it is no longer periodic. More importantly, it lowers the barrier between the periodic wells, so that the lower energy levels can "feel" the top of the barrier (non-parabolic), and the qubit has a large anharmonicity even if $E_c \ll E_J$ (Fig. 6). Therefore, with a large shunt capacitor, a phase qubit is insensitive to charge noise but still has a large anharmonicity. Its first and second lowest energy levels will then be used to represent state $|0\rangle$ and state $|1\rangle$, respectively. It can also be seen in Fig. 7 that the eigenstates have a small spread of phase. So they have relatively well-defined phases, and thus, the qubit is called a phase qubit. However, a phase qubit is sensitive to $1/f$ flux noise, which will affect the bias current through wire inductances [79]. Moreover, the bias current is usually near $I_c$ and state $|1\rangle$ is near the top of the barrier. Tunneling of state $|1\rangle$ out of the potential well is easy (which is also the mechanism for a readout process in a phase qubit, see Fig. 7). Therefore, $T_1$ is usually low ($< 1\mu s$) [80]. Since it has a small loop inductance, $E_c \ll E_L$.

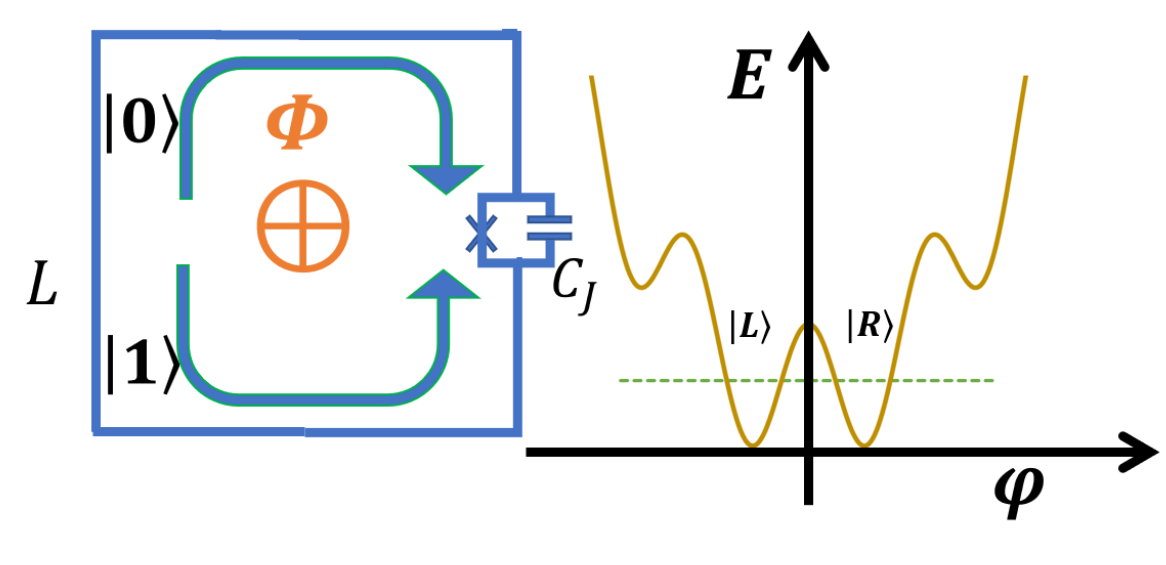

FIGURE 8. A flux qubit is biased with a half magnetic flux quantum. Its energy landscape is modified to have a double well due to the external flux and the loop inductance $L$. The superposition of the states in the left and right wells give rise to clockwise and counterclockwise persistent currents, respectively, which are used to encode |0⟩ and |1⟩.

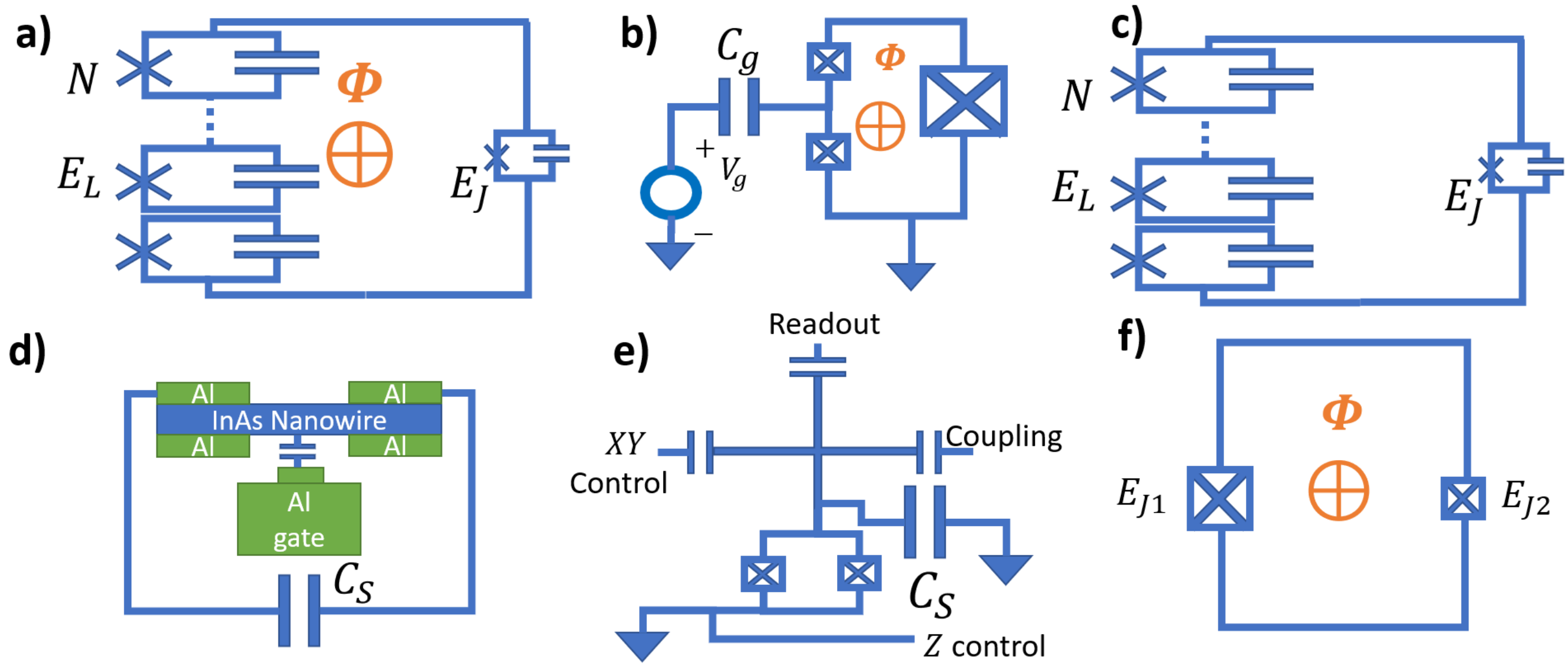

**FIGURE 9. Other types of qubits. a) Fluxonium: It has a series of large-area Josephson junctions to form a superinductor to avoid large parasitic capacitance. b) Quantronium: It needs an external flux and gate biases to work at the double sweet spot. c) Blochnium: It is similar to Fluxonium but does not require an external flux bias. d) Gatemon: In this example, its Josephson junction is formed by Al/InAs/Al stack. The charge density in InAs nanowire can be turned by a gate to change the critical current of the Josephson junction. The figure shows a transmon because it is shunted by a large capacitor. e) Xmon: Xmon refers to its layout. It has a pair of Josephson junctions forming a loop to allow flux tuning by the Z control line. The Z control line flows current to create the necessary flux. It has a large shunt capacitor. Therefore, it is still a transmon. One of the capacitor plates is laid out as a cross (X) to allow capacitive coupling to the readout circuit, XY control line, and qubit-to-qubit couplers. f) Flux-tunable qubit: It is implemented with a superconducting quantum interference device (DC SQUID) loop. Its effective Josephson inductance can be changed by an external flux.**

Another type of superconducting qubit is the flux qubit [77], [81] (Fig. 8). The simplest flux qubit is obtained by replacing the current source with a wire in a phase qubit with an external magnetic flux threading through the loop. It also does not have a large shunt capacitor. Its Hamiltonian is given by [77],

$$H = 4E_c n^2 - E_J \cos\varphi + \frac{(\Phi_0\varphi/2\pi - \tilde{\phi})^2}{2L}, \quad (14)$$

where $\tilde{\phi}$ and $L$ are the external flux and the loop inductance, respectively. The last term is due to the flux-induced energy in the loop. When $\tilde{\phi} = \Phi_0/2$ (i.e., half of the magnetic flux quantum), the potential energy (sum of the second and third terms) of the system has two double wells, symmetric about $\varphi = \pi$, as shown in Fig. 8. They are degenerate, similar to a double dot quantum well. The wavefunction of the state in the phase space can be at the lowest level on the left well ($|L\rangle$) or the right well ($|R\rangle$). Due to tunneling through the barrier, the degeneracy is lifted, and the eigenstates become a bonding and an anti-bonding one. They are used as the qubit states with $|0\rangle = (|L\rangle + |R\rangle)/\sqrt{2}$ and $|1\rangle = (|L\rangle - |R\rangle)/\sqrt{2}$ (cf. charge qubit at the sweet spot). It turns out that $|0\rangle$ and $|1\rangle$ physically correspond to a loop current circulating the loop clockwise and anti-clockwise. Therefore, in a flux qubit, a general state is a superposition of clockwise and anticlockwise circulating currents. It should be noted that for a better tunneling coupling between $|L\rangle$ and $|R\rangle$ and to increase the anharmonicity, flux qubits are usually formed by 3 Josephson junctions [82].

A flux qubit has $E_c \ll E_J$ [83]. This is achieved by having a large Josephson junction instead of an external shunt capacitor. For example, its critical current is 500 nA in [83] compared to the charge qubit with $I_c < 50\ nA$ in [84], which has $E_c \gg E_J$. Therefore, it is insensitive to charge noise, but it is sensitive to flux noise, which can influence the third term in Eq. (14). As a result, it has only moderate $T_1$ (e.g., $12\mu s$ in [81] and $25\mu s$ in [85]) and $T_2^*$ (e.g., $2.5\mu s$ [81]).

To reduce flux noise sensitivity, one may use a large shunt inductor so that $E_J \gg E_L$ [25]. However, it is difficult to realize a large inductor using a superconducting wire. This is because a geometric wire with a large inductance also has a large parasitic capacitance. Therefore, fluxonium was proposed in [86]. A fluxonium can be seen as an improvement of a flux qubit by replacing the loop geometric inductor with a superinductor formed by a large number (e.g., 43 in [86]) of large-area Josephson junctions in series, instead of a large geometric inductor (Fig. 9a). Due to the series connection, the total inductance $L$ is still larger than that of the original Josephson junction inductance, $L_J$ (so $E_J \gg E_L$). The superinductor also does not have a large parasitic capacitance. In another view, the superinductor reduces its sensitivity to flux noise by modifying the energy landscape in phase space so that the wavefunction is more spread out in the phase space and the qubit frequency is less sensitive to flux variation at sweet spots. It should be noted that flux noise is about 100 times smaller than charge noise in nature [25].

There are two other advantages of fluxonium. It has a lower qubit frequency (<1 GHz) than other types of qubits (>3 GHz). It is known that dielectric loss reduces with qubit frequency [87] and, therefore, its decoherence time suffers less from dielectric loss (Section VIII). Moreover, it has a large anharmonicity in the order of GHz, which allows the application of short pulses. While the early development of fluxonium only had a moderate $T_1$ (e.g., $10\mu s$) and $T_2^*$ (e.g., $2\mu s$) [88], in recent years, high coherence times with $T_1$ and $T_2^*$ exceeding 1 ms have been achieved [89], [90], [91]. However, one should note that qubit coherence times may degrade substantially in a quantum circuit. Very often, $T_1$ and $T_2^*$ are only tens of $\mu s$ [87].

### C. Other Types of Superconducting Qubits

There are other types of superconducting qubits. We will discuss a few of them briefly.

Quantronium is based on CPB. It has a loop by splitting the Josephson junction and has a large Josephson junction for readout (Fig. 9b) [92]. It operates at the sweet spots of both charge and flux to be first-order insensitive to both charge and flux noise with $E_C \sim E_J$. Therefore, neither $\varphi$ nor $n$ is maximally localized. As a result, the sensitivities to flux and charge noise are balanced.

Borrowing the idea of reducing charge sensitivity in a transmon qubit by using a large shunt capacitor, we may use a large shunt inductor to make it flux insensitive. Blochnium, which is mathematically a dual of a transmon, is thus proposed in [93] (Fig. 9c). It has $E_c/E_L \gg 100$ and $E_c \sim E_J$. With these parameters, the potential in the phase space acts like a crystal lattice and, thus, crystal bands are formed. Therefore, its states are Bloch states (thus the name Blochnium). Like fluxonium, the large inductance is usually achieved by using a superinductance. And external flux is not needed for a Blochnium.

Another type is gatemon [94], [95], [96], which stands for "gate tunable transmon". A gatemon has a Josephson junction formed by two superconductors (e.g., Al, NbTiN, etc.) separated by a semiconductor instead of an insulator (Fig. 9d). So it is abbreviated as S-N-S, with N for normal. Very often, a gatemon is shunted by a capacitor to form a transmon. The semiconductor is usually a nanowire (e.g., InAs, carbon nanotube). The weak link between the superconductors depends on the current density in the semiconductor, which can be controlled electrostatically by an external gate. Since the weak link determines $I_c$, the gate voltage can be used to adjust the Josephson energy and qubit frequency (Eq. (9), with a note that while the S-N-S Josephson junction still shows Josephson effects, the current phase relationship in Eq. (5) is usually non-sinusoidal [97]). However, since its qubit frequency is controlled by the gate voltage, it is more susceptible to charge noise compared to a regular transmon. It should also be noted that gatemon is not the same as gmon, which is a type of coupler to be discussed in Section VI.

A cat-qubit (or Schrödinger cat qubit) encodes a quantum state using coherent states [98], [99]. A coherent state is a minimum-uncertainty quantum state whose time evolution follows classical motion [14], [100]. They can be made to be resilient to a certain type of error and thus minimize the requirement of quantum error correction. For example, in [99], the cat-qubit is resilient to bit-flip error and has achieved more than 10s of bit-flip time.

Another commonly mentioned qubit is called Xmon [101], [102] (Fig. 9e). However, it is named based on its layout, and it is usually just a transmon.

If two Josephson junctions are formed in a loop, their effective Josephson energy can be tuned by an external magnetic flux [14]. This is based on a superconducting quantum interference device (DC SQUID) loop. When the two junctions are identical (with $E_J$), under a total flux of $\Phi$, the total Josephson energy $E_{JT}$ is

$$E_{JT} = 2E_J \cos\left(\frac{\pi\Phi}{\Phi_0}\right) \quad (15)$$

This effective Josephson junction can then be used to form different types of qubits mentioned earlier. Fig. 9f shows a flux-tunable qubit.

## VI. QUBIT COUPLING

The main reason to enable qubit-qubit coupling is to implement two-qubit gates (Criterion 4). The most relevant two-qubit gates are entanglement gates (such as $CNOT$, $CZ$ (controlled-phase gate), $iSWAP$, $\sqrt{iSWAP}$), which are essential to form a universal set of quantum gates [14], [26] as mentioned in Section II. Single-qubit gates have reached over 99.9% fidelity and are above the fault-tolerant threshold for some fault-tolerant algorithms. However, two-qubit gates are still the major bottleneck for efficient fault-tolerant quantum computing. Particularly, in some cases, it can reach 99.9% in an isolated system, but it reduces to less than 99% in a non-isolated system. The major source of infidelity comes from parasitic coupling to the next nearest neighbor (NNN) qubit, leakage to non-computational space, and crosstalk [103]. There are various types and classifications of coupling schemes. Often, they can be distinguished by whether the qubits are brought to resonance, or whether the coupler is tunable, or whether the process is all-microwave-drive, or whether the process is AC or DC driven, etc. We will discuss various types of coupling schemes below and refer to these characteristics whenever appropriate. We will also concentrate on the discussion of two-qubit gates for basic qubits such as charge qubits, phase qubits, flux qubits, and transmons.

### A. Fixed Coupling with Tunable Qubits

One may couple two fixed qubits through a fixed coupler. However, the coupling strength is constant, and the frequency of one qubit will be dependent on the state of the other qubit, as the coupling is always on. To avoid this problem, two qubits with different frequencies can be coupled through a fixed capacitor or mutual inductance. When coupling (execution of a two-qubit gate) is needed, the qubits will be brought to resonance, which means that the energy levels of certain states of the system will be made to be degenerate or near-degenerate. Capacitive coupling is used for phase and charge qubits, and inductive coupling is used for flux qubits (Fig. 10).

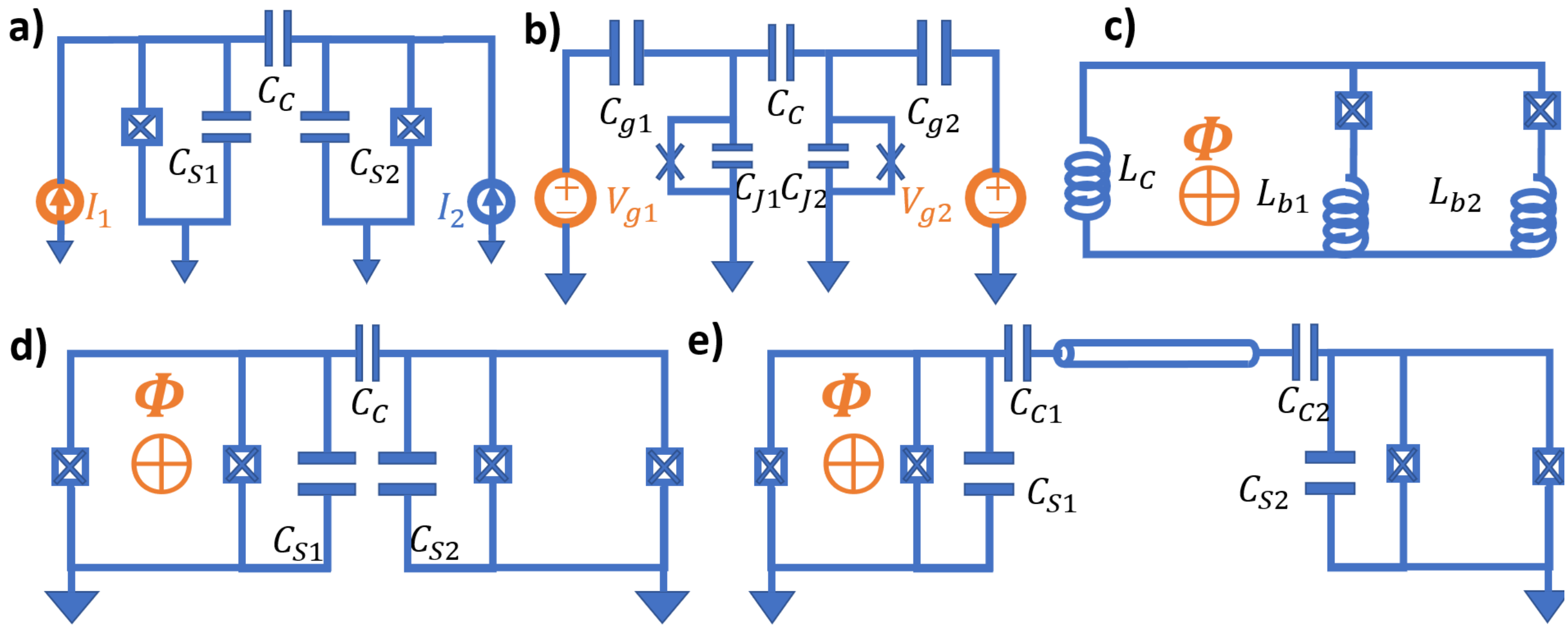

**FIGURE 10.** **Various two-qubit coupling schemes. The tunable parameters are highlighted in orange. a) Phase qubits are capacitively coupled through $C_c$. Bias current $I_1$ is changed to enable the $CZ$ gate. b) Charge qubits are capacitively coupled through $C_c$. Bias voltages $V_{g1}$ and $V_{g2}$ are changed to enable the $CNOT$ gate. c) Flux qubits are inductively coupled through $L_c$. Bias flux in the common loop is changed to enable controlled operations. Note that a flux qubit usually has 3 junctions. Only one is used here for simplicity. d) Flux-tunable transmon qubits are capacitively coupled through $C_c$. Bias flux in one of the qubits is changed to enable the $iSWAP$ gate. e) Flux-tunable transmon qubits are coupled through a quarter-wavelength transmission line. Bias flux in one of the qubits is changed to enable coupling.**

For phase qubits, one can change the qubit state energy levels by changing the biasing current to enable two-qubit SWAP-like and $CZ$ gates [104], [105] (Fig. 10a). In the implementation of a $CZ$ gate, non-computational states $|02\rangle$ and $|20\rangle$, which primarily interact with $|11\rangle$, are borrowed to accumulate an extra phase on $|11\rangle$ by bringing $|11\rangle$ to be near-degenerate with $|02\rangle/|20\rangle$.

For charge qubits, one may use the offset charge to change the energy manifold to enable resonance between selected states (Fig. 10b). In [106], gate pulses are applied to the two qubits so that the offset charges on the two qubits are changed and $|00\rangle$ and $|01\rangle$ are brought into degeneracy, with the first qubit as the control qubit. This enables the coupling between $|00\rangle$ and $|01\rangle$. If the gate pulse is applied for the right amount of time, it will swap $|00\rangle$ and $|01\rangle$. On the other hand, the gate pulse amplitude is not large enough to bring $|10\rangle$ and $|11\rangle$ into degeneracy. Therefore, it acts as a conditional gate that flips the target qubit when the control qubit is $|0\rangle$.

Inductive coupling is usually used for flux qubit coupling. The mutual inductance can be formed through various methods, such as sharing a common leg of the two qubit loops [107] or sharing a common inductance [108]. In [108], microwave-field-induced flux through the common loop is used to implement a 2-qubit gate (Fig. 10c). The transition matrix due to the external flux is such that, at the $\tilde{\phi} = \Phi_0/2$ bias point, only the transition between two levels is allowed (non-zero element due to selection rules), enabling controlled two-qubit operations.

Similarly, transmon qubits can also be directly capacitively coupled to implement $iSWAP$ gate [14] or $CZ$ gate [102], [109]. In the $iSWAP$ gate case, the qubits are flux-tunable and brought to the same frequency during the 2-qubit operation for a suitable amount of time (Fig. 10d). For $CZ$ gate case, the flux-tunable qubits are tuned to bring $|11\rangle$ to be near-degenerate, forming an avoid-crossing, with $|02\rangle/|20\rangle$ for an extra phase accumulation, like in the charge qubit case. The extra phase accumulation in $|11\rangle$ is due to its energy shift under the interaction with $|02\rangle/|20\rangle$ energy levels.

There is a general trade-off between gate time and error rate. When the gate time is short, it is less affected by the decoherence time of the individual qubit. However, there will be more high-frequency components in a short gate pulse, which will enable undesirable transitions (e.g., leakage from computational space to non-computational space), such as $|11\rangle$ to $|02\rangle/|20\rangle$ in $CZ$ gate. Therefore, careful pulse engineering is required. For example, adiabatic CZ gates with reduced leakage but a relatively fast gate time have been proposed with a projected error less than $10^{-4}$ for transmon qubits [110].

Another way to further reduce the two-qubit gate time with minimum leakage to the non-computational space is to optimize the coupling constant and qubit non-linearity such that the intended swapping ($|01\rangle$ and $|10\rangle$ in an $iSWAP$-like gate) and the undesirable swapping ($|11\rangle$ to $|20\rangle$ leakage) are synchronized. As a result, a diabatic two-qubit gate as fast as 18 ns and a Pauli error rate as low as $4.3\times10^{-3}$ has been achieved [111].

To couple qubits far from each other, a quantum bus can be used (Fig. 10e). One of the implementations is to use a superconducting transmission line resonator as a cavity [112]. The flux-tunable transmon qubits are capacitively coupled to the ends of the half-wavelength resonator (at voltage antinodes). When the two qubits are brought to the same frequency, the exchange of their states (swapping of $|01\rangle$ and $|10\rangle$) is mediated by virtual photons in the cavity and can be used to implement the $\sqrt{iSWAP}$ gate. More formally, we can add the photon state of the cavity at the end of the notation of the total system (i.e., $|qubit1, qubit2, photon\rangle$ ) to describe the process. Then, the interaction between $|010\rangle$ and $|100\rangle$ is achieved via the exchange of a virtual photon in the cavity, with the intermediate state being $|001\rangle$. Therefore, it is operated in the dispersive qubit-cavity regime to avoid cavity-induced loss. It is dispersive because $|001\rangle$ has a different energy than $|010\rangle$ and $|100\rangle$ (detuned). The photon is virtual because there is no real photon occupation in the cavity.

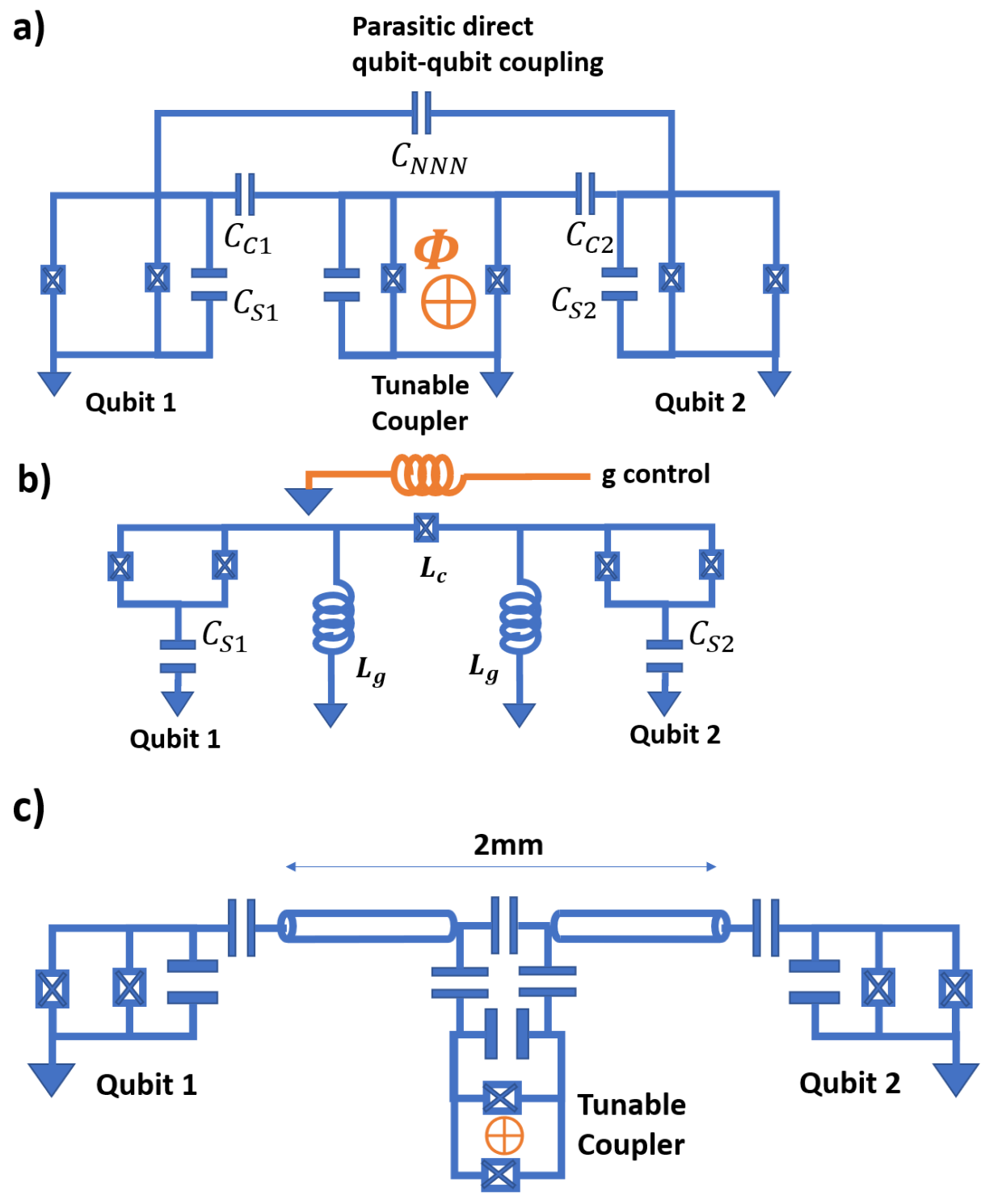

**FIGURE 11. Tunable coupler examples. The tunable parameters are highlighted in orange. a) An example of qubit-qubit coupling through a tunable coupler. The two qubits are transmon qubits. They are coupled to a tunable coupler whose frequency can be tuned by an external flux (which essentially is a flux-tunable transmon qubit). There is also a direct qubit-qubit coupling through parasitic capacitors. NNN refers to next-nearest neighbor. b) gmon coupler. Note that the transmon qubits have a small linear inductor ($L_g$) to close the loop. The coupling inductor is a Josephson junction coupled to an inductor biased by a g-control line. c) Tunable coupler mediated by transmission lines to allow long-distance coupling.**

Similarly, the qubits can also be coupled to a lumped LC oscillator. Like the transmission line, the lumped LC oscillator acts as a cavity [113].

### *B. Tunable Coupling*

In a tunable qubit scheme (fixed coupler), the qubit frequency needs to be swept to enable a two-qubit gate operation. This increases the chance of hitting defect frequencies, resulting in decoherence (see two-level-system defects in Section VIII). Moreover, the bias line for qubit frequency tuning can introduce noise. This can be mitigated with a tunable coupler. The idea of having a tunable coupler is to turn on and off the coupling during and outside of a two-qubit gate operation, respectively. Thus, the qubits can operate at their optimal points. A tunable coupler can mediate the interaction between the two qubits through virtual excitations (such as a photon). Therefore, it is not reliant on qubit-qubit resonance, and thus the qubits may have a larger frequency difference (as there is no need to tune their frequencies) to reduce cross-talk and allow a larger scale of integration. Tuning is achieved, for example, by adjusting the coupler's resonant frequency [103]. Unlike fixed couplers, the interaction can be minimized when a two-qubit gate is not in action. Therefore, the larger coupling strength can be used and the gate time can be reduced from hundreds of ns in a fixed coupler to tens of ns in a tunable coupler. This provides a better gate-time-to-fidelity trade-off. However, this increases the circuit complexity, and the tunable coupler may also introduce a new decoherence channel and crosstalk for the qubits.

Fig. 11a shows the scheme of a tunable coupler. The difference between the coupler and qubit frequency (detuning), $\Delta$ is much larger than the coupling strength, $g_i$ (with $i = 1{,}2$), so that it is in the dispersive regime to suppress qubit leakage to the coupler. It should be noted that there is also undesirable finite direct coupling between the two qubits, $g_{12}$ (due to $C_{NNN}$). Luckily, the effective coupling between the two qubits (direct qubit-qubit and through the coupler) is,

$$\tilde{g} = \frac{g_1 g_2}{\Delta} + g_{12}. \quad (16)$$

Since $\Delta < 0$, $\tilde{g}$ can be tuned to 0 to completely turn off the coupling [103]. This can offset the direct qubit-qubit (NNN) coupling. It is predicted that a gate fidelity higher than 99.999% can be achieved in 100 ns in the absence of decoherence.

There are different ways to make a tunable couple. One is to use a flux-tunable qubit to form a variable LC-tank [103] (Fig. 11a). The coupling can be flux-tuned to modulate qubit-qubit coupling by operating the coupler in the dispersive regime through virtual exchange. Through careful engineering, an $iSWAP$ gate with a 30 ns gate time and 99.87% fidelity has been achieved for transmon qubits through a Josephson junction LC-tank tunable coupler [114].

Another famous coupler is called the "gmon" coupler (not to be confused with gatemon in Section V) [115]. It eliminates DC coupling between the qubits to avoid crosstalk (Fig. 11b). Their connection is interrupted by a Josephson junction, which blocks DC signals. An adiabatic $CZ$ gate has been demonstrated. The two qubits (transmons) are connected to ground through their respective small inductors ($L_J \gg L_g$, to maintain the nonlinearity) and coupled to each other through a Josephson junction, which acts as a flux-tunable coupling inductor $L_C$, which is tuned the flux generated by the current going through a g-control line. An excitation current in one qubit will flow to the small inductor attached to the other qubit and generate a magnetic flux to affect the state of the other qubit. The amount of current depends on the coupler inductance.

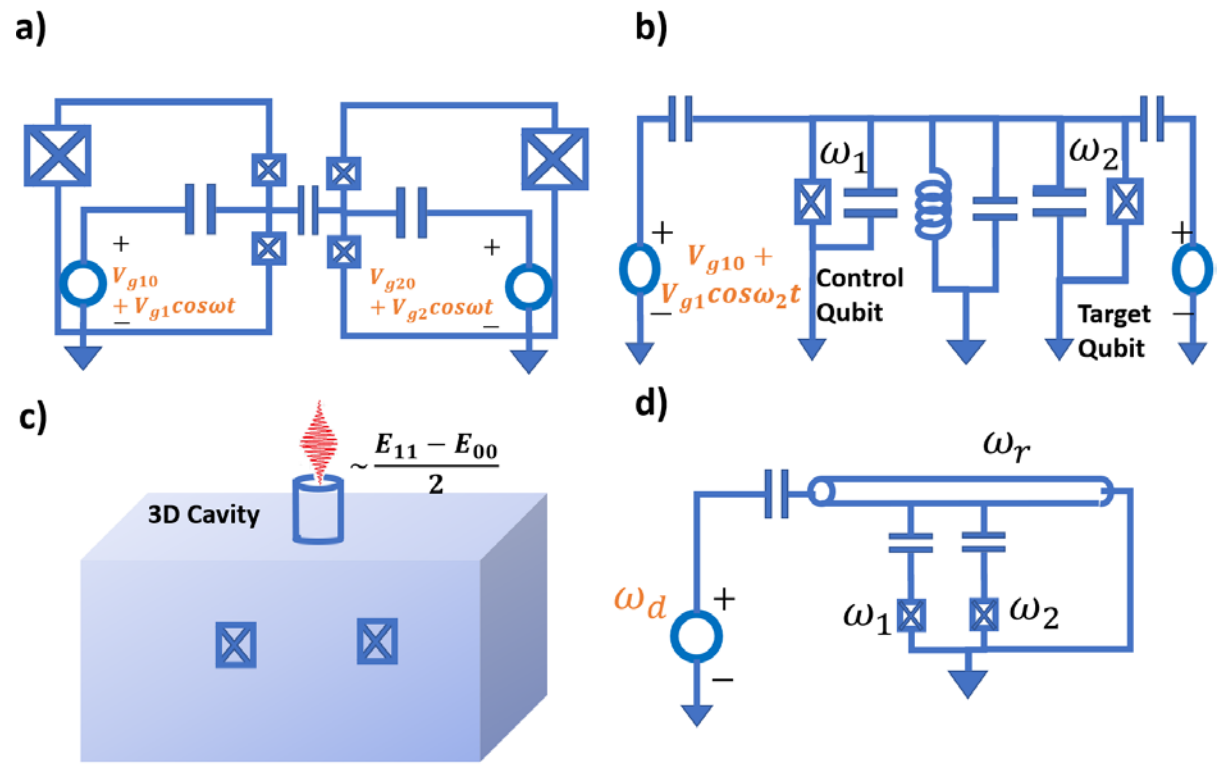

**FIGURE 12. All-microwave drive coupling examples. The microwave drive signals and sources are highlighted. a) An example of a phase qubit all-microwave drive coupling with a fixed coupling capacitor. b) Cross-resonance coupling of two transmon qubits through a fixed cavity (LC tank). Qubit 1 is driven at the qubit frequency of qubit 2. c) Two transmon qubis are coupled through a 3D resonator with a suitable microwave drive. d) Resonator-induced-phase shift gate for two transmons. The two transmons are coupled to a resonator, which is driven by a microwave with a suitable frequency, amplitude, and duration.**

To allow long-distance coupling, a tunable floating-transmon coupler, mediated by waveguides between the components, instead of capacitive coupling, has been demonstrated [116] (Fig. 11c). The two qubits are separated by 2 mm. This has the potential to reduce cross-talk and allow more layout flexibility. A $CZ$ gate with fidelity > 99.8% has been achieved.

### *C. Fixed Coupling with All Microwave Drive*

While tunable couplers may provide a better gate-time-to-fidelity trade-off over fixed couplers, the complex circuit control may hinder the implementation of large-scale quantum computers. And as discussed, fixed couplers require the ability to tune the qubit frequency during gate operation. This might result in coupling to unwanted non-computational space and spurious environmental mode [117]. Therefore, an all-microwave drive two-qubit gate using a fixed coupler is desirable. In [118], the quantronium qubits are coupled by a fixed capacitor, and each qubit is driven by an off-resonant microwave pulse through the gate (Fig. 12a). The effective Hamiltonian is then equivalent to two qubits with a tunable coupler. As a result, entanglement can be created.

Another type of all microwave drive scheme is the cross-resonance CR gates (Fig. 12b). CR gates were proposed in which, during a two-qubit gate operation, the control qubit is driven at the target qubit's frequency [119]. This does not need flux tuning during the 2-qubit gate operations and is all microwave-based control. For example, in [117], capacitively shunted flux qubits (CSFQ) coupled via a fixed resonator are driven through AC flux due to an AC flux bias current to achieve a $CNOT$ gate. To further increase gate fidelity, echoed cross-resonance (ECR) is proposed by adding an echo pulse to remove unwanted components in the Hamiltonian [120]. It is then further improved through various engineering techniques [121], [122], [123], [124] and is being used on one of the major commercial superconducting quantum computer platforms.

Another implementation of an all-microwave drive two-qubit gate uses a 3D cavity to house two detuned qubits dispersively coupled to the cavity [125] (Fig. 12c). A microwave with a frequency approximately half of the energy difference between $|00\rangle$ and $|11\rangle$ is applied to the system via the cavity. This allows $|00\rangle$ to $|11\rangle$ transition (which was forbidden) through two-photon interactions. This process is enhanced by making the $|0\rangle$ to $|1\rangle$ transition of one qubit approaches that of $|1\rangle$ to $|2\rangle$ of the other, and by interacting with higher-level non-computational states. The use of a high-quality factor 3D cavity also enhances the process. This implements a $bSWAP$ gate which mixes $|00\rangle$ and $|11\rangle$. When it is applied to the ground state $|00\rangle$, it creates a Bell state.

The resonator-induced-phase (RIP) gate is another type of all-microwave drive gate (Fig. 12d) [126]. RIP gates can couple two qubits even if they are far detuned from each other. Therefore, there is less constraint on the qubit frequency, which favors large-scale integration. The two qubits are coupled to a bus resonator dispersively. The resonator frequency is shifted depending on the qubit state (cross-Kerr, see Section VII). Therefore, through an off-resonant drive of the resonator, the two qubits acquire a state-dependent phase to implement a $CZ$ gate. In [127], a theoretical study shows that a gate time of 120 ns with an infidelity of $6\times10^{-4}$ can be achieved. In [128], it is experimentally demonstrated in a 4-qubit system coupled to a 3D cavity, with the control-target qubit frequency detuning spanning from 380 MHz to 1.8 GHz. In [129], *ab initio* analysis of the RIP gate dynamics was conducted and showed that less anharmonicity and higher qubit frequencies reduce leakage.

One may also combine CR and RIP settings to increase gate fidelity. In [130], a fast 40ns $CNOT$ gate with error less than $10^{-4}$ is predicted.

Other types of all-microwave 2-qubit gates include resonator sideband-induced $iSWAP$ gate [131], microwave-activated controlled-PHASE (MAP) gate [132], cross-cross resonance (CCR) operation for $iSWAP$ gate [133], etc.

### *D. Parametric Tunable Coupler*

A parametric tunable coupler is similar to the tunable coupler mentioned earlier. However, instead of tuning the coupler with a DC or slow AC flux, it uses AC flux on top of a DC bias to achieve time-dependent modulation of a parameter such as the capacitance or inductance of the coupler (Fig. 13) [134]. In other words, a regular tunable coupler provides static coupling controls. In a parametric tunable coupler, the coupling strength is also tuned during the gate time to increase and customize interactions. The interaction between the qubits is resonantly activated. This allows the implementation of a faster gate.

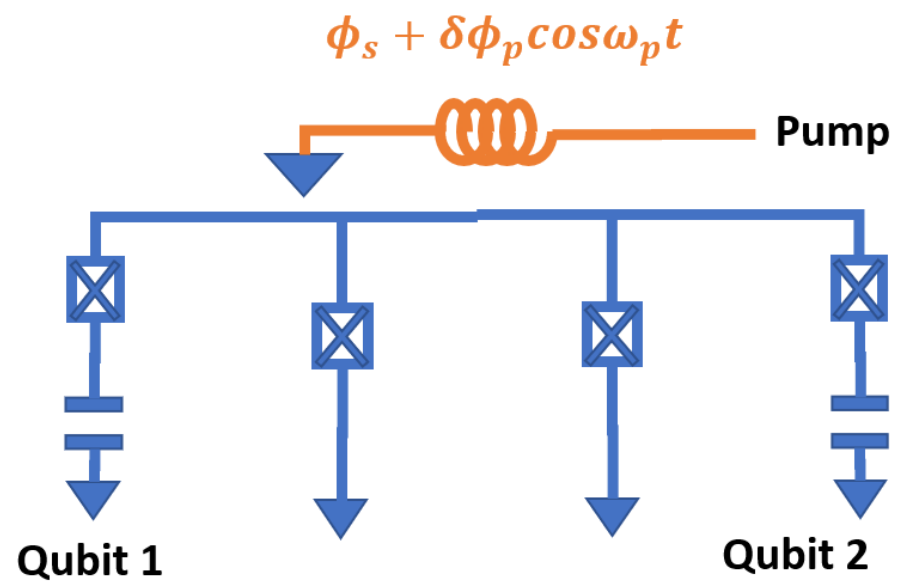

FIGURE 13. **Example of a parametric tunable coupler. Two transmon qubits are coupled to each other through the middle Josephson junction loop. The flux in the loop is modified by the pump signal. It is instructive to compare this to gmon in Fig. 11b.**

For example, in [134], two transmon qubits share a two-junction parametric coupler (which is just a flux-tunable qubit). An AC flux through the coupler is driven by the pump circuit. The coupler inductance is modulated at the pump frequency, so is the coupling strength. This generates additional coupling between two of the two-qubit states. It is shown that by changing the pump frequency, the coupling can change from a parametrically-induced resonant type to a dispersive type. Therefore, it can perform both $iSWAP$-type and $CZ$-type gates. Compared to a non-parametric tunable coupler, it is faster (30ns with 99.47% fidelity for a $CZ$-gate). However, it is more flux sensitive and more difficult to calibrate.

## VII. READOUT

As mentioned in Section II and Fig. 2, dispersive readout is commonly used to read out the state of a transmon qubit (Criterion 5) [135]. In this paper, we will only discuss this mechanism in depth.

In dispersive readout, a resonator is coupled to a qubit, and the resonant frequency, $\omega_r$, of the resonator is modified by the qubit state. If the qubit frequency, $\omega_q$, is far detuned from the resonator (large $|\Delta| = |\omega_q - \omega_r|$ compared to their coupling strength, $g$), the shift of the frequency, cross-Kerr $\chi$, is given by [14], [28],

$$\chi = -g^2 \frac{E_C/\hbar}{\Delta(\Delta - E_C/\hbar)} \tag{17}$$

where $g$ and $\hbar$ are the resonator-qubit coupling strength and reduced Planck constant, respectively. Note that here $g$ is in frequency. In some literature, it is the coupling energy [14]. Such a measurement scheme is nearly quantum nondemolition (QND) at low power levels (a few photons). In a QND measurement, the collapsed eigenstate of the measurement observable would not be changed after every measurement.

One of the major design parameters in qubit readout is the trade-off between readout speed and qubit decay. To enable a fast readout, the coupling between the qubit/resonator and the environment needs to be strong. However, this also means that the decay rate of the qubit will increase.

For a large-scale integration, multiple qubits share the same feedline. In some early literature, two qubits were coupled to the same resonator, and the resonator frequency shift is based on the cross-Kerr effect due to the joint state of the two qubits [136]. It was also demonstrated by simulation that single-shot readouts can be performed for two qubits when they have their own individual resonators with a small frequency spacing [137]. But in more recent experiments, each qubit has its own resonator [138], [139], and the readout can be performed through frequency multiplexing.

The readout measurement can be multiple-shot or single-shot. In a multiple-shot measurement, an identical experiment needs to be run multiple times. If the measurement is QND, it can be measured directly multiple times to reduce error without rerunning the algorithm. A longer readout pulse, in general, increases the readout fidelity. However, if the measurement time (readout pulse) is too long, qubit decoherence might have changed the qubit state during the measurement, regardless of the measurement scheme. Therefore, while long readout time (pulse) can reduce noise, it can increase errors due to relaxation from $|1\rangle$ to $|0\rangle$ and repopulation from $|0\rangle$ to $|1\rangle$, too [139]. For example, in [138], the re-thermalization error and mixing of $|0\rangle$ to $|1\rangle$ due to readout tone accounts for 0.3-1.2% of readout error, 0.7% due to decay, and 0.5% due to noise. Therefore, it is important for accurate characterization of readout error, including readout-induced leakage rate to non-computational space [140].

Very often, single-shot measurement is preferred as it provides short turnaround, which is important for conditioning quantum-state initialization, error correction, feedforward circuits, and quantum network [138], [141]. To achieve single-shot measurement, a stronger qubit-resonator coupling is required. This can decrease its coherence time. Therefore, a single-shot measurement usually needs a Purcell filter. Moreover, it does not allow noise averaging as in multiple-shot measurement, and a quantum-limited amplifier (QLA) is needed (Fig. 1). It should be noted that there are other schemes to allow single-shot measurement without using QLA, which will not be discussed [142].

In the following subsections, we will discuss Purcell filters and QLAs.

### A. Purcell Filters

When a quantum system (e.g., a qubit at $\omega_q$) is coupled to a resonator or cavity (at $\omega_r$) with a coupling strength of $g$, the spontaneous emission of the quantum system will be modified. The emission (decay) rate is given by [143], [144]

$$\Gamma_{Purcell} \approx \kappa \left(\frac{g}{\Delta}\right)^2 \propto \frac{Re[Y(\omega_q)]}{C_q}, \tag{18}$$

where $\Delta = \omega_q - \omega_r$, $\kappa$, $Y(\omega_q)$, and $C_q$ are the qubit-cavity detuning, resonator line width, admittance seen by the qubit, and qubit capacitance, respectively. This is called the Purcell effect. It is a result of the interaction between the qubit and the outside world through the resonator. When the coupling is strong (large $g$) and the line width is wide (large κ) to increase the readout speed (and fast reset), qubit decay is also fast and, thus, has a lower $T_1$. The Purcell filter is introduced in [144] to solve the dilemma to protect qubit from spontaneous emission while maintaining a strong coupling to a low-Q cavity. In Fig. 14, the Purcell filter shorts the transmission line to ground at the qubit frequency so that it has zero real admittance to shut off the decay channel. In [139], it is shown that with a Purcell filter, the decay rate can be reduced to

$$\Gamma_{Purcell} \propto \kappa \left(\frac{g}{\Delta}\right)^2 \left(\frac{\omega_q}{\omega_r}\right) \left(\frac{\omega_r/Q_F}{2\Delta}\right)^2, \qquad (19)$$

where $Q_F$ is the quality factor of the filter (e.g., 30). Similar to [144] and Fig. 14, a quarter-wave (λ/4) coplanar waveguide resonator is used as a Purcell filter in [139], but it is embedded into the feedline. It has four qubits. Each has its own resonator but shares the same feedline. Due to the low $Q_F$, the passband is wide enough to allow the reading of four qubits through multiplexing. The Purcell filter is shared by all qubits. All reach 99% fidelity with a pulse shorter than 300 ns.

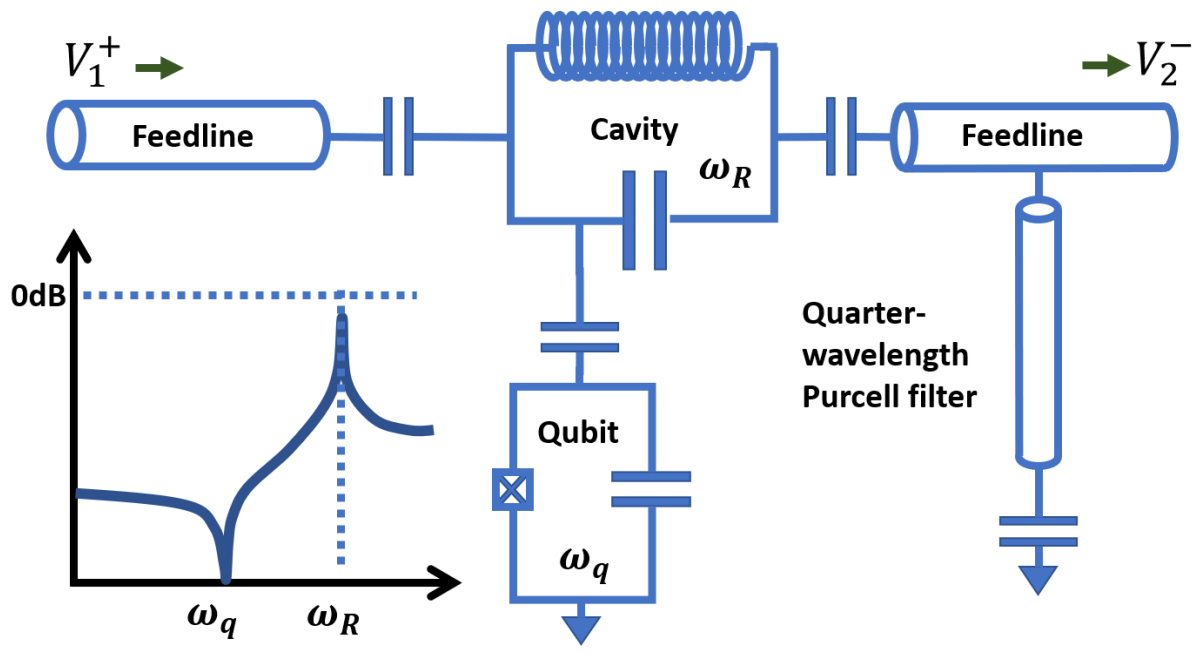

**FIGURE 14.** **A dispersive readout circuit with a Purcell filter (implemented using a quarter-wavelength transmission line). The readout pulse is injected from the left and measured at the right. Like in Figs. 1 and 2, the transmission power is maximum only near the cavity's resonant frequency (peak at $\omega_R$). With the Purcell filter, an extra dip occurs at the qubit frequency, $\omega_q$, because the quarter-wavelength transmission line shorts the feedline at $\omega_q$. This avoids the qubit decaying through the cavity.**

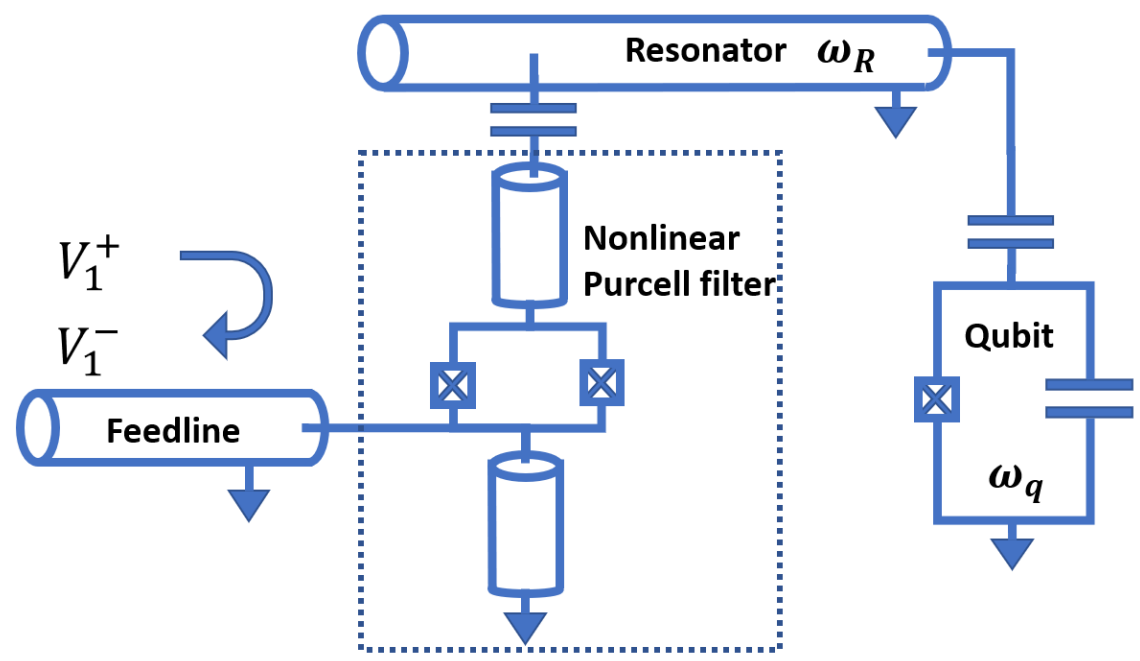

**FIGURE 16.** **Non-linear Purcell filter used in a readout circuit. The readout is performed by measuring the reflection instead of transmission. A nonlinear Purcell filter is formed by a SQUID interrupting a quarter-wavelength resonator. The filter is inserted between the resonator and the feedline.**

Purcell filters become more important as the readout fidelity is improved. This is because the readout error is limited by the qubit decay during measurement [33]. In [33], measurement time is reduced by co-optimizing various parameters (increasing the dispersive interaction, choosing optimal readout resonator linewidth, employing a Purcell filter, and parametric amplification). It achieved a readout 98.25% accuracy in less than 48 ns or 99.25% accuracy for 88 ns.

In a multiple-qubit system, besides sharing one Purcell filter (e.g., [139]), one may also assign Purcell filters to individual resonators (Fig. 15). For example, in [138], a 5-qubit multiplexed readout was demonstrated, and each resonator has its own Purcell filter. Besides protecting qubits from Purcell decay, it also minimized off-resonant driving in untargeted readout resonators during the readout process (i.e., avoiding affecting qubit 1 when reading qubit 2). As a result, an average of 97% correct assignment was achieved for the 5 qubits with an 80 ns readout pulse.

Residual photons in the resonator during readout can cause dephasing in the qubit because the qubit frequency depends on the number of photons in the resonator [145]. A non-linear Purcell filter has been proposed to suppress the dephasing without sacrificing the readout speed [146] (Fig. 16). It is non-linear because its line-width and transmission coefficient depend on the amplitude. As a result, dephasing is suppressed during idle time (low photons) and enhanced during readout (high photon counts). A λ/4 resonator interrupted by a superconducting quantum interference device (SQUID) is used to implement the non-linear filter between the readout resonator and the readout waveguide. A fidelity of 99.2% with a 40 ns readout pulse has been achieved.

Even with a non-linear Purcell filter, the energy relaxation time is still limited [146]. Therefore, an additional filtering mechanism is introduced to further enhance the coherence time by engineering an "intrinsic" filter. An "intrinsic" filter can also save space to enable better integration. In [32], this is implemented using a 3D cavity. This paper utilizes the distributed element and multi-mode nature of the resonator to reduce decay. The coupling port to the resonator is at the voltage node in the dressed-qubit mode. As a result, the decay rate is low. On the other hand, at the resonator fundamental mode, it is away from the voltage node and thus has sufficient coupling for fast readout (Fig. 17). In [147], an intrinsic filter is achieved by creating a notch filter through the coupling of a λ/4 readout resonator to a dedicated λ/4 filter resonator. The notch filter broadens the linewidth of the readout mode. The coupling region creates destructive interference at the qubit frequency, resulting in a notch filter to suppress qubit decay. 99.9% fidelity was achieved with a sub-60 ns pulse.

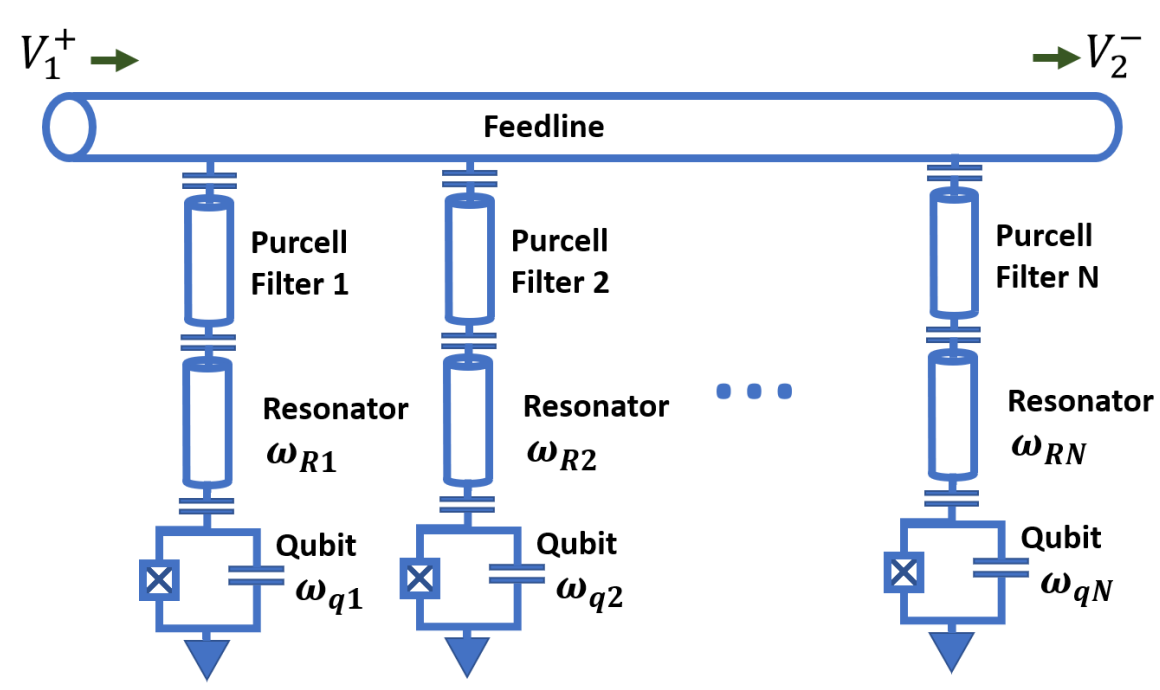

**FIGURE 15.** **A multiple-qubit system sharing the same feedline. Each resonator has its own Purcell filter. This not only reduces Purcell decay of the qubits but also avoids driving non-targeted qubits during the readout of the target qubit.**

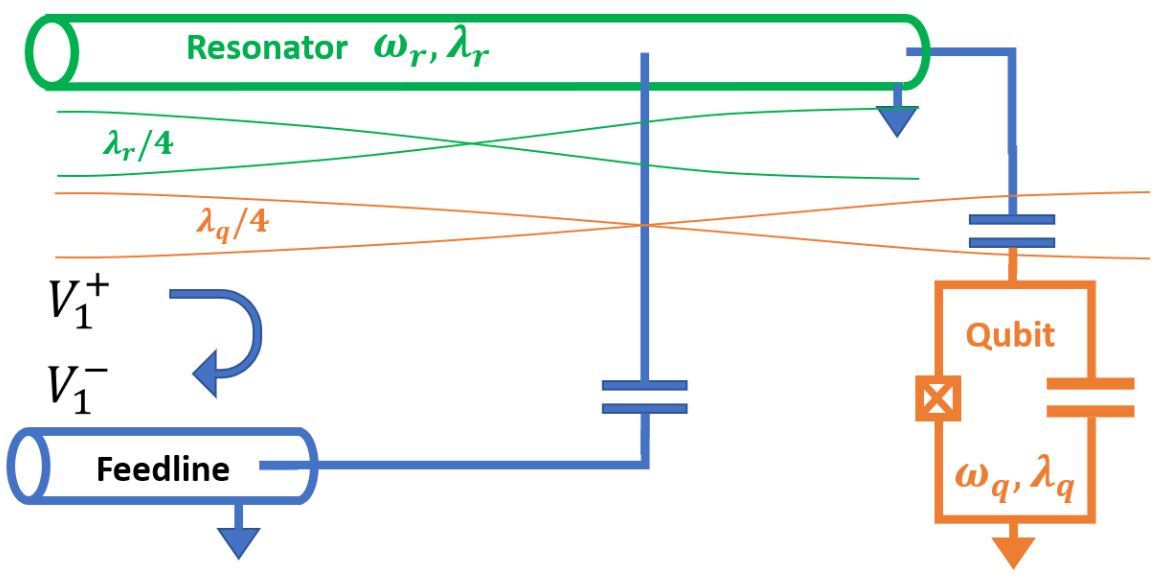

FIGURE 17. Idea of an intrinsic Purcell filter. The transmission line resonator is a half-wavelength resonator with both sides open (so maximum in voltage). The readout feedline is coupled to a point in the resonator such that it is the voltage node for the qubit mode (orange) and thus has a minimal interaction to reduce qubit decay. On the other hand, it is away from the voltage node of the resonator mode (green) so that it has a strong coupling to the resonator to enable fast reading (i.e., allow the resonator photon to leak into the feedline).

To further avoid the area impact on the qubit chip due to the incorporation of Purcell filters, in [148], a 3D reentry cavity is used to couple to the qubit chip capacitively. This also allows post-fabrication adjustment of coupling strength. It works as a large-linewidth bandpass filter with intrinsic Purcell filtering.

And to further reduce qubit decay during readout, in [149], by adjusting the detuning between the qubit and the readout resonator dynamically, the readout speed is improved without sacrificing other metrics. During readout, a fast flux pulse is applied to change the qubit frequency so that it has a smaller detuning with the resonator frequency. Therefore, the dispersive shift is larger during readout, which will improve the readout fidelity and speed. At its idle time, the qubit has a large detuning from the resonator and thus has a low decay rate. It achieved an error rate of $2.5\times10^{-3}$ for a 100 ns pulse.

### *B. Quantum Limited Amplifiers*

In a readout process, amplification is required so that the weak photon signals can be detected. However, as mentioned in Section II, every amplifier amplifies both the signal and noise but also adds noise to its output. The corresponding metric is the noise figure (NF). As a result, the signal-to-noise ratio is degraded after each stage of amplification (larger NF has more degradation). Due to Friis formula (Eq. (20)), it is desirable to have an amplifier with the lowest NF at the first stage. This is because the total noise figure of an $N$-stage amplification circuit, $NF_{Total}$, is given by,

$$NF_{Total} = 1 + (NF_1 - 1) + \frac{NF_2 - 1}{A_{P1}} + \frac{NF_2 - 1}{A_{P1}A_{P2}} + \cdots, \quad (20)$$

where $A_{Pi}$ and $\mathrm{NF}_i$ are the available power gain and noise figure of stage $i$ for $i = 1$ to $N$, respectively. Therefore, $NF_1$ dominates while the noise figures of other stages are scaled down by the gains in the previous stages.

Quantum-limited amplifiers (QLAs) have the lowest NF in theory and are usually needed for single-shot measurement [150]. In the best case, it adds half a photon of noise in the amplification process.

QLAs are usually put at the beginning of an amplification chain, as shown in Fig. 1. A desirable QLA should have a low NF, be directional, have a large gain, have a large gain compression limit, and have a large bandwidth. A QLA is usually implemented by parametric amplification. A detailed review of quantum-limited parametric amplifiers is given in [151]. Parametric amplification utilizes the non-linearity of a circuit. Therefore, a Josephson junction, which has non-linear inductance, is a natural choice. A parametric amplifier takes a pump source, $\omega_p$, to amplify the signal, $\omega_s$. The outputs are the amplified signal and an idler $\omega_i$ (Fig. 18).

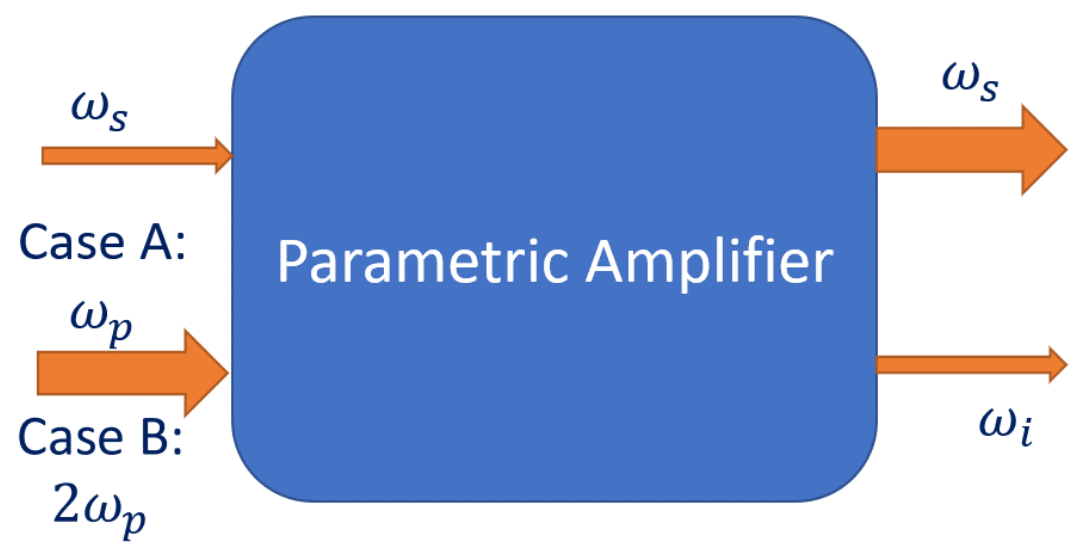

FIGURE 18. Block diagram of a parametric amplifier. It requires a pump ($\omega_p$) to amplify a signal ($\omega_s$). Two examples are given. For case A, one pump photon is converted into one signal photon and one idler photon ($\omega_i$) with $\omega_p = \omega_s + \omega_i$. For case B, two pump photons are converted to one signal photon and one idler photon with $2\omega_p = \omega_s + \omega_i$.

Parametric amplifiers can be classified based on degeneracy, the number of ports, order of wave mixing, etc.

In terms of degeneracy, they can be classified as degenerate and non-degenerate parametric amplifiers. In a degenerate parametric amplifier, the idler and signal are identical (within the respective linewidth). The phase difference between the pump and the signal determines the energy flow and amplification. Therefore, it is phase-sensitive. It can noiselessly amplify one quadrature of a small signal (the noise in the other quadrature will exceed the quantum limit due to the uncertainty principle). In a non-degenerate parametric amplifier, the gain does not depend on the phase and, thus, it is phase-insensitive. The idler will adjust itself accordingly.

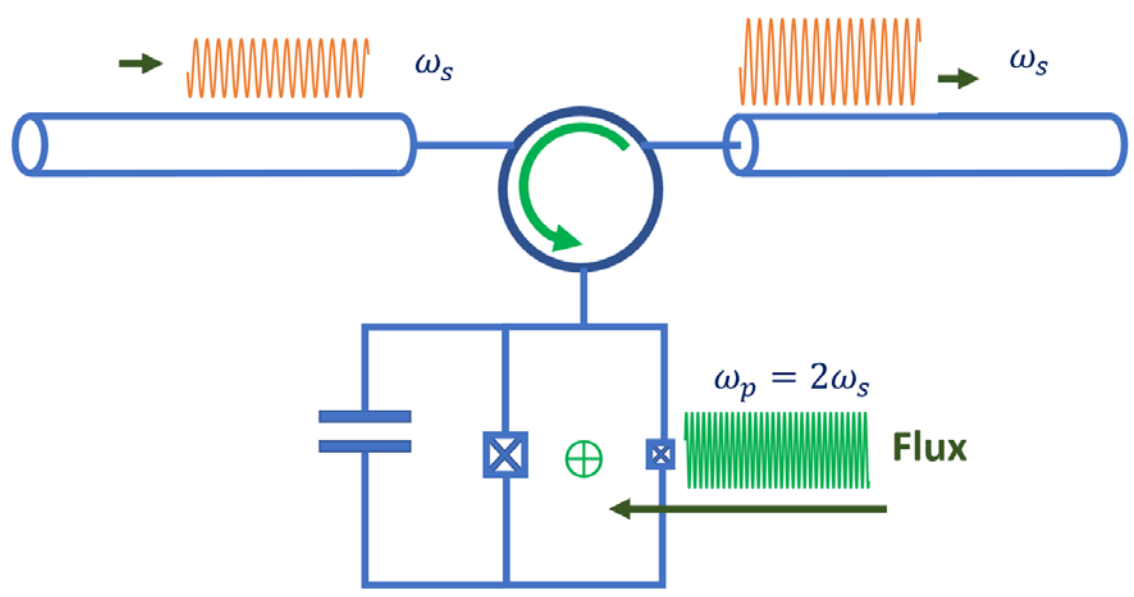

**FIGURE 19.** Degenerate three-wave mixing using JPA. A flux as the pump is applied to a SQUID structure at $2\omega_s$. The pump modifies the Hamiltonian of the JPA. Through the circulator, the input signal is amplified in the JPA through three-wave mixing.

In terms of the order of wave mixing, they use either four-wave (4WM) or three-wave mixing (3WM). Four-wave mixing uses the third order of nonlinearity, which naturally occurs in Josephson junctions, and does not need symmetry breaking. An example process is $2\omega_p = \omega_s + \omega_i$ [152], [153]. Three-wave mixing uses the second order of nonlinearity. An example process is $\omega_p = \omega_s + \omega_i$ [154]. Three-wave mixing is only possible if the system does not have inversion symmetry. Therefore, in degenerate cases, $\omega_p = \omega_s$ in 4WM and $\omega_p = 2\omega_s$ for 3WM.

Josephson parametric amplifier (JPA) and traveling wave parametric amplifier (TWPA) are the most commonly used parametric amplifiers.

A JPA is formed by a Josephson junction-based resonator (Fig. 19). It is a cavity/resonator with anharmonicity due to a Josephson junction to allow wave mixing through non-linearity. A JPA can detect a few photons and has a 10-50 MHz gain bandwidth with 20 dB gain [155], [156]. In [157], 20dB with 640MHz bandwidth was achieved by engineering the imaginary part of the environment impedance.

TWPA uses a long chain of Josephson junctions as a metamaterial transmission line [152] (Fig. 20). Therefore, in principle, it is also a JPA because it uses Josephson junctions for parametric amplification. However, it is usually distinguished from JPA as JPA is usually referred to the lumped circuit one. It has a wider bandwidth and can reach GHz (e.g., 20dB over 3 GHz in [152]). It can also handle a higher power. Therefore, it is suitable for multi-qubit readout through frequency multiplexing [152].

Besides JPA and TWPA, there are other types of parametric amplifiers that try to achieve better figures of merit. For example, to avoid the bulky isolator after the QLA (see Fig. 1, where the isolator is used to avoid backward noise propagation), a traveling-wave parametric amplifier and converter (TWPAC) has been proposed [158]. Through phase-matching, it is a TWPA in the forward direction to amplify the signal, and it is a frequency converter in the backward direction to up-convert and down-convert any back-propagating noise out of the signal band. This may reduce the hardware overhead for larger-scale integration.

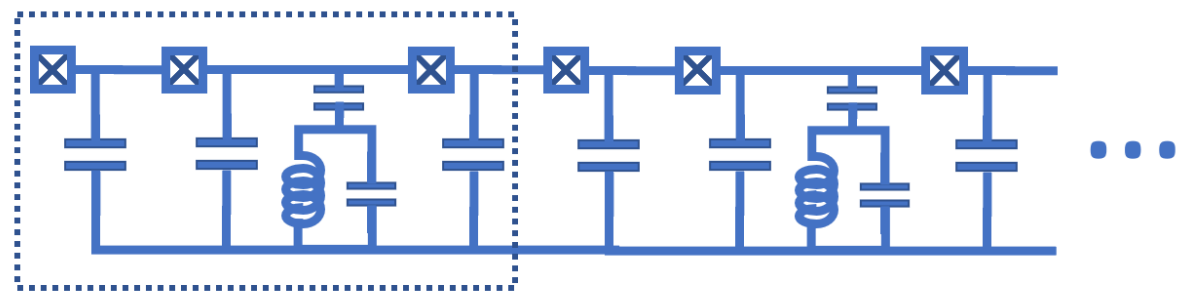

**FIGURE 20.** Schematic of a JTWPA. Two unit cells are shown. It can be compared to the telegrapher's equation model of a transmission line, where inductors are replaced by Josephson junctions. Additionally, an LC-resonator is inserted in each unit cell.

The superconducting nonlinear asymmetric inductive element (SNAIL) parametric amplifier (SPA) is introduced in [159]. A SNAIL contains $n$ large Josephson junction and a small Josephson junction in a loop, threaded by an external flux. This is similar to a flux qubit under certain parameters. But a SNAIL operates in a regime to keep the cubic term and cancel the fourth-order term of the Hamiltonian, with the flux breaking the symmetry. Thus, it has a non-zero second-order nonlinearity, and three-wave mixing is possible. This also allows the freedom to further engineer and improve the compression power [160]. Since it is biased with a DC flux, this is also a type of DC-biased amplifier.

Another type of parametric amplifier is the kinetic inductance traveling wave amplifiers [154], [161]. It uses the non-linear kinetic inductance of superconducting film for parametric amplification. Suitable thin films are niobium nitride (NbN), titanium nitride (TiN), molybdenum-rhenium (MoRe), and niobium titanium nitride (NbTiN). It can be used to make a phase-sensitive parametric amplifier.

One type of parametric amplifier that may be promising is the Floquet-mode traveling-wave parametric amplifier [162]. A Floquet mode is a time-periodic eigenstate of a periodically driven Hamiltonian, analogous to an eigenstate of a static system. Floquet-mode amplifiers prevent information leakage, provide a wide bandwidth and high quantum efficiency at the same time, and suppress forward-backward wave coupling. It is shown that a gain of > 20 dB and a near-ideal quantum efficiency can be achieved.

## VIII. TWO-LEVEL SYSTEM DEFECTS

While there are many possible loss mechanisms in superconducting qubits (such as loss due to two-level system (TLS) defects, dielectric loss in Josephson junction oxide, and loss due to non-equilibrium quasi-particles [144]), TLS defects are believed to be the primary barrier to further improving the superconducting qubit coherence time (Criterion 3). It also affects the resonant frequency and quality factor of superconducting resonators. A detailed review of TLS can be found in [163].

The physical origin of TLS is unclear. It may be due to tunneling atoms, tunneling electrons, spin and magnetic impurities, or other origins [163]. Among them, tunneling of atoms between two potential minima is the most appealing one. In disordered materials such as amorphous oxide, an atom or a small group of atoms may tunnel between two different configurations [164] (Fig. 21). Atomistic tunneling is particularly important at low temperatures because the thermionic emission above the tunneling barrier is impossible; and at higher temperatures, both the lower and higher energy states are equally populated, TLS defects can no longer absorb energy (saturated) and has less observable effect. This is described by the standard tunneling model (STM). In this tunneling process, the atoms move in the order of one interatomic distance. Indeed, in [165], it is found that the dipole distance of the TLS agrees with the expected value of about 0.13nm, which is about an atomic bond length. The energy difference between the two configurations results in two energy levels. When the qubit or resonator has its frequency close to that of the TLS, energy loss occurs through the interaction between the TLS dipole and the electric field. Due to the random nature of disordered materials, TLS frequencies are widely distributed and, thus, it is difficult to avoid.

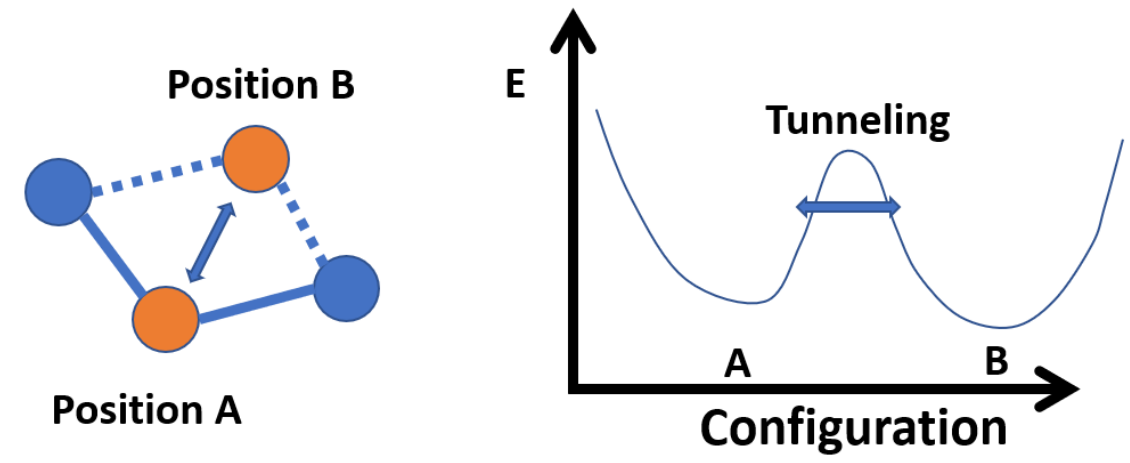

**FIGURE 21.** **Left: Illustration of an orange atom switching between positions A and B in real space. Blue atoms are assumed to be fixed. Right: The energy landscape in the configuration space in the standard tunneling model. The switching is achieved by the orange atom tunneling through an energy barrier between positions A and B.**

There are four common locations of TLS. They are the substrate-air interface, metal-air interface, substrate-metal interface, and Josephson junction insulator (Fig. 22). The first three interfaces are sometimes called the circuit interface [166]. There has been substantial evidence that TLS defects do not exist in the bulk of the material. Fig. 22 shows a typical layout of a transmon qubit, and the white areas are the substrate-air interfaces.

The interaction between a qubit and TLS is strong when the electric field at the TLS defects is large. In [101], Xmon's with the same capacitance and junction parameters but different gap sizes between the capacitive plates are studied. When the gap size is large, the electric field is small as the potential is dropped across a larger distance. It is found that the Xmon $T_1$ time increases with the gap size, which is consistent with the TLS theory. It is also deduced that most of the interaction occurs at 100 nm from the metal edge, where the electric field is the largest, which interacts with a sparse bath of incoherent defects. Similar qualitative observations are also found in [167]. If a 3D cavity is used, due to its larger volume, the electric field is more spread out, and it is also less sensitive to surface dielectric. As a result, the effect of TLS defects can be reduced [168].

To understand the source of TLS defects, electric field spectroscopy is used to distinguish defects due to Josephson junction oxide and circuit interface [166]. In the technology used in [166], it is found that circuit interface TLS contributes 60% of the dielectric loss. Moreover, while loss due to Josephson junction oxide accounts for 40%, the loss is mostly from the large area parasitic one due to the shadowing manufacturing method (see Fig. 3a). Sources of TLS include photoresist residual, impurity atoms, and substrate damage.

Therefore, to reduce the impact of TLS by Josephson junction oxide, one may also reduce the junction area to avoid tunnel defects [165], [167]. To reduce the TLS on the circuit interface, a better insulator process should be used. In [165], when CVD $SiO_2$ is replaced by $SiN_x$, the phase qubit's coherence time is increased by 20 times because OH in $SiO_2$ is the major source of TLS in that process. As mentioned in

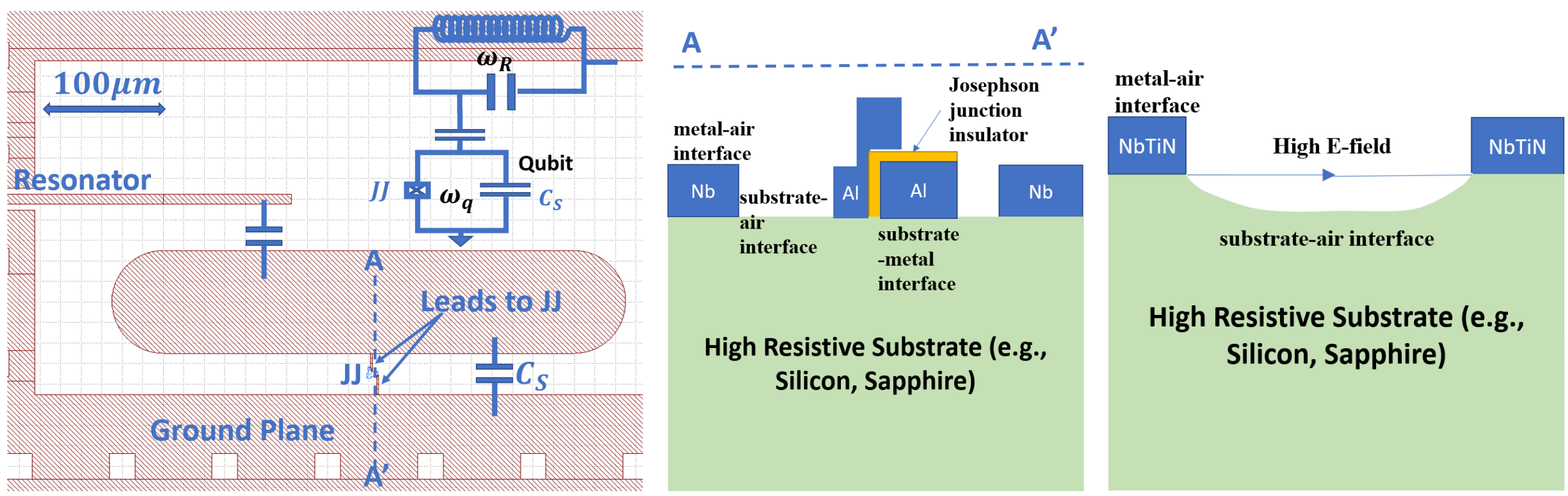

**FIGURE 22.** **Left: An example layout of a transmon qubit capacitively coupled to a readout resonator. Only the coupling part of the resonator is shown. The corresponding circuit (middle part of Fig. 14) is shown as the inset. The white areas are exposed high-resistive substrates (> 10,000 Ω·cm$^2$, e.g., Si, Sapphire, also see Fig. 3). Pink areas are superconducting niobium at cryogenic temperatures. Middle: Cross-sectional view of AA'. TLS defects are believed to exist in four types of locations, as indicated. They are the metal-air interface, substrate-air interface, substrate-metal interface, and the Josephson junction tunneling barrier (insulator). Right: Etching of substrate to avoid the interaction of substrate-air interface TLS defects with the high electric field between the metals.**

Section IV, crystalline tunneling barriers have also been studied in order to minimize the impact of TLS found in amorphous barriers (such as rhenium(Re)/crystal-$Al_2O_3$/Al Josephson junction [64] and epitaxial growth of NbN/AlN/NbN junction [65]).

To minimize the impact of photo-resist on decoherence time, one may also modify the process to minimize the area patterned by shadowing. In [167], an extra step is used to pattern the leads (see example of leads in Fig. 22, left) of the Josephson junction with a regular lithograph, and only the region close to the junction is patterned by liftoff (short-liftoff). Compared to the structure with the whole lead patterned by liftoff (long-liftoff), short-liftoff has a larger $T_1$. Moreover, it is found that, in the statistical distribution of $T_1$, long tail of low $T_1$ is observed in both short-liftoff and long-liftoff, which is due to the edge defect within 500 nm of the junction. Statistical study of $T_1$ is crucial for large-scale integration.

The understanding of qubit-TLS interaction and the properties of TLS is also further confirmed with the resonator-TLS interaction. The resonant frequency of a resonator will shift when it interacts with TLS because TLS changes the effective dielectric constant. The shift is proportional to the ratio between the energy stored in the TLS-loaded volume and the total energy in the resonator volume [169].

In [169], a niobium resonator is used. The experiment shows that the TLS defects are distributed within a few nanometers from the surface. Temperature-dependent shift of resonator frequency is also observed, which confirms that the defects are distributed on the surface of the CPW due to native oxide.

It is found that the loss of the co-planar waveguide in millikelvin is material independent [170]. This strengthens the belief that TLS lies on the interface and surface. In [171], by removing substrate from the high E-field region, the quality factor is increased for a NbTiN resonator (Fig. 22, right).

Therefore, with proper cleaning and a high-quality crystalline substrate, a high-quality factor (~$10^7$) resonator was achieved in [172]. In that experiment, the single-crystalline sapphire is used as the substrate. Al, which has a robust oxide, is used as the metal. The substrate was cleaned thoroughly without being damaged before the molecular beam epitaxial (MBE) Al deposition. These reduce the substrate-metal, substrate-air, and metal-air defects, resulting in high quality resonator.

A study to distinguish the dielectric loss and mechanical loss (coupling of phonon to TLS) in amorphous silicon also confirmed the previous understanding of TLS. In [173], it is found that they do not have the same origin. Dielectric loss is due to the coupling of the electric field to dangling bonds, but mechanical loss correlates with density. Mechanical quality factor can be increased by orders of magnitude when $SiO_2$ is grown at higher temperature, but not much for dielectric loss.

To further understand the nature of TLS, the coupling between TLS with resonator under off-resonant drive was studied in [174]. This was used to characterize non-linear properties and dephasing rate, and identify the contribution of TLS to the total loss.

Recently, using scanning gate microscopy, the microscopic nature of TLS near a NbN resonator at millikelvin temperatures has been further understood [34]. The TLS is "visualized" and its dipole orientation was extracted, which can be used to understand the origin and nature of TLS.

## IX. ELECTRONIC DESIGN AUTOMATION

In this section, we attempt to parallel the electron design automation (EDA) in superconducting qubit chip design with that in semiconductor chip design. EDA is expected to be indispensable for the large-scale integration of superconducting qubit quantum computers. Establishing an EDA ecosystem built on the current one, spanning from atomistic simulation to system design, is necessary [175].

### *A. Atomistic and Microscopic Calculations*

In semiconductors, *ab initio* or atomistic simulations play a critical role in modern transistor designs. This includes calculating the band structure of devices with small volumes and under stress [176], calculating defect levels and formation energies [177], understanding novel materials [178] and carrier transport [179], among other applications. In superconducting qubits, *ab initio* calculations are used in many aspects. One of the most prominent ones is the understanding and calculation of two-level system (TLS) defects, which are believed to be the major limitation of the coherence time (see Section VIII). *Ab initio* calculation can be used to calculate the formation energy of certain types of defects under various fabrication conditions (e.g., O-rich or Al-rich in alumina growth) and then calculate the defect properties (e.g., dipole, potential barrier) to identify TLS of a frequency that is deleterious to qubit coherence time [180]. It can be used to confirm and explain the observations in the experiment [181] and then propose new fabrication methods. For example, in [182], calculations are performed to confirm that H atoms are a source of flux noise due to their spin, and coating the source with graphene is proposed to eliminate the noise source. It can also be used to simulate the fabrication process to understand the junction uniformity and reproducibility [183].

Monte Carlo simulations are used in semiconductors to understand non-equilibrium transport, implantation, and epitaxial growth by taking parameters extracted from *ab initio* calculations [184]. Similarly, Monte Carlo simulations are commonly used to understand stochastic processes in a superconducting qubit. For example, it is used to understand the interaction of TLS and a qubit [165].

### *B. Finite Element and Macroscopic Simulations*

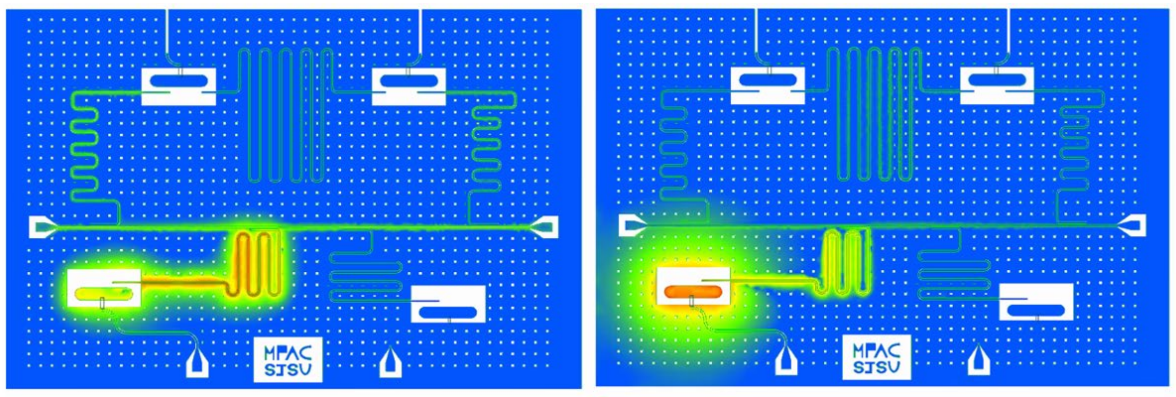

**FIGURE 23. Eigenmode simulation of a qubit chip using ADS QuantumPro. Left: Resonator mode. Right: Qubit mode. The junction is replaced by a linear inductor, $L_J$, in this classical simulation. The energy distribution can be used in the EPR method to calculate anharmonicity and cross-Kerr.**

In semiconductors, Technology Computer-Aided Design (TCAD), which solves the governing equations of carrier transport and the Poisson equation, is an essential tool to design new devices and understand device behaviors [185], [186]. In superconducting qubit systems, various simulation approaches are used to assist the design of the superconducting qubit system.

Electromagnetic (EM) Simulation: EM simulations are used to design microwave circuits (such as resonators, feedlines, Purcell filters), qubits, and quantum amplifiers. Qubit manipulations and readout circuit performance depend on the coupling capacitors and inductors, which in turn determine the quality factor of the qubits and thus the maximum qubit coherence time. The quality factor of the readout resonator determines the readout speed. Therefore, EM simulations have been used substantially for quantum circuit design [14]. More importantly, for 3D integrations, finite element simulations can provide deep insight into the cross-talk between various layers [187], [188]. In some tool modules, non-linear effects are included to simulate kinetic inductance, which is important for the design of parametric amplifiers [189]. But it should be noted that, very often, the Josephson junction is only replaced by an equivalent L-C circuit in the simulation [187] (Fig. 23). Common commercial tools are Ansys High Frequency Structure Simulator (HFSS) [187], COMSOL [147], Keysight Advanced Design System (ADS) [189], and QTCAD [190].

Qubit Design Frameworks: In EM simulations, classical quantities are extracted. They are then used to calculate qubit properties such as qubit frequency, qubit anharmonicity, and the cross-Kerr among the qubits and the resonators. There are two major methods for performing these calculations, namely the black-box quantization (BBQ) [191] and the energy participation ratio (EPR) [192]. In the BBQ method, it approximates the Josephson junction potential into a linear and a non-linear part. The linear part contains the parallel linear capacitor and linear inductor, and then they are combined with the linear circuit elements (such as the cavity). The linear part is transformed into a sum of parallel RLC resonators through Foster's theorem. Each resonator represents a mode. This whole circuit is then made in parallel with the nonlinear part of the Josephson junction. Finally, the circuit is quantized [14] with the non-linear part as a perturbation.

Therefore, in BBQ, finite-element simulations are used to extract the impedance of the circuit, which can then be used to calculate the cross-Kerr of each mode. The BBQ method accounts for the coupling between a multi-level qubit and a multi-mode cavity by solving the coupled linearized problem and incorporating the weak anharmonicity of the Josephson junction. This is applicable in the transmon regime where charge dispersion is negligible.

In the EPR method, it also separates the system Hamiltonian into a linear part and a nonlinear part, with the nonlinear part as the perturbation. However, the non-linear coefficient in the circuit is achieved by finding the so-called energy participation ratio (EPR) of each mode in each circuit element. This can be calculated by finding the electric and magnetic field distributions through eigenmode analysis in finite-element simulations (Fig. 23). The ratios can then be used to calculate the cross-Kerr and anharmonicity.

Qubit Dynamics Simulations: After extracting various classical and quantum mechanical properties, depending on the type of qubits, the coupling energy can be extracted to construct the Hamiltonian. The Hamiltonian can then be fed into the master equations to perform qubit dynamics simulations. One of the common platforms is QuTip [193], which is an open-source software for simulating the dynamics of open quantum systems. This can be used to, for example, simulate two-qubit gates [114], [129], [130], [194].

### *C. Circuit Description and Schematics to Layout*

In semiconductor analog circuit design, it is often necessary to design a circuit using a schematic, which will then be converted to a layout. Simulations are performed before and after the layout (pre-layout vs. post-layout simulations). For digital design, hardware description language (HDL) can be converted to a netlist and eventually a layout.

In some tools, one may design a quantum circuit using schematics and perform circuit simulations. For example, in [195], SQUIDs, tunable couplers, and parametric amplifiers are simulated after schematic circuits are constructed.

In superconducting qubit circuit design, one may also describe the circuit through Python language, which will then create the corresponding layout (Fig. 24). One of the commonly used platforms is Qiskit Metal [196]. It can also interface with a finite-element simulator to perform EM simulations to obtain the necessary quantities for EPR or BBQ extractions.

### *D. Co-Simulations*

Nowadays, parasitic resistance and capacitance become the bottleneck in semiconductors. Therefore, design-technology co-optimization (DTCO) [185] and system-technology co-optimization (STCO) [197] are essential in the development of advanced technology nodes. Given that a qubit is very sensitive to the environment and noise, that manipulation and readout pulses are analog in nature, that cross-talk is a limiting factor to fidelity, and that gate speed and coherence time contradict each other, it is expected that co-simulations and co-optimizations are important.

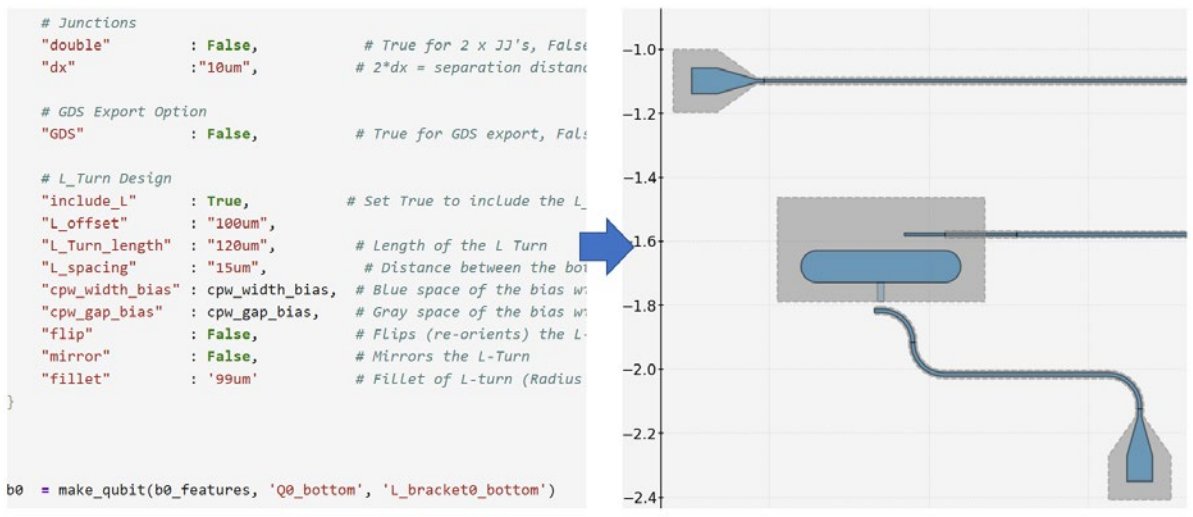

**FIGURE 24. Creation of a superconducting qubit circuit layout (right) using Python code (left) in Qiskit Metal.**

For example, one may perform readout pulse optimization to achieve the best fidelity-pulse-width trade-off by using a compact model with parameters extracted from finite-element simulations and taking the noise generated in the readout chain into consideration [198]. Machine learning can also be used for optimization for this problem [199]. A behavior model has also been proposed to simulate control electronics to obtain their manipulation waveform [200]. It can be used to understand how digital precision can change the fidelity of a quantum gate.

Since co-simulation involves both classical and quantum components and also involves small-scale and large-scale components, sometimes appropriate transformation and modeling are needed. In [201], a systematic methodology is used to transform the quantum components to circuit models to enable co-simulations with classical control electronics.

## X. LARGE-SCALE INTEGRATION

### *A. Overview*

The success of the large-scale integration of a superconducting qubit quantum computer requires advancement in various technologies and the establishment of a healthy ecosystem [202]. Here, we try to list some of the development directions to enable the large-scale integration of superconducting qubit quantum computers.

1. Coherence and TLS Defects: We need to continue to improve the qubit coherence time (Criterion 3). This is because the improvement of physical qubit coherence time will reduce the number of physical qubits needed to form a high-quality logical qubit. Instead of relying on the 60-year-old lift-off method, novel qubit fabrication (such as in [61]) and wafer cleaning methods need to be continued to be developed. For example, it is believed that ion milling before bottom electrode deposition leaves about 3.9 nm of interfacial layer underneath the metal in a lift-off process (metal-substrate interface), which has a loss tangent in the order of $10^{-3}$. A bandage process is, thus, proposed to protect the substrate to alleviate the problem [203]. Bridge-free processes have also been proposed (e.g., [204]). We also need to have a better understanding of TLS through simulation and experiment.
2. Qubit Density and Heterogeneous Integration: Innovative ideas to increase the qubit density without deteriorating qubit crosstalk, gate fidelity, and readout fidelity are needed. Examples are 3D integration and microwave frequency multiplexing [139]. In 3D integration, flip-chip technology can be used. Signal and readout circuits can be put on one chip, and the qubits can be put on another [205]. They can be indium bump-bonded or placed on a carrier chip. Matured CMOS technologies, such as Silicon interposer [206] and through silicon via (TSV) [207] would be used. 3D integration also allows heterogeneous integration. In this way, various types of qubits and various circuit components can be optimized separately [25]. Methods to reduce circuit size, such as obviating the needs of preamplifiers [208] and Purcell filters (using intrinsic Purcell filters), are needed.
3. Long Distance Quantum Interconnect: A modular design with high-fidelity inter-chip quantum connections is needed for large-scale integration. An efficient transducer needs to be developed to allow inter-cryostat communications [209]. For example, a nonlinear microwave resonator can be used (e.g., lithium niobate ($LiNbO_3$) or aluminum nitride (AlN)) [210].
4. Dilution Refrigerator: Currently, control signals are generated by high-speed classical circuits at room temperature [211], [212]. Without multiplexing, each qubit requires at least three cables (one for the control signal, one for readout, and one for flux tuning). While a superconducting qubit quantum computer does not generate heat, each cable conducts about 0.013 μW of heat power from room temperature to the mixing chamber where the qubits are hosted at ~20 mK [213]. A typical dilution refrigerator has a cooling power of less than 50 μW [214]. This means that only about 4,000 cables can be connected, which translates to approximately 1,000 qubits. Note that cables for other purposes, such as those used in quantum-limited amplifiers and tunable couplers, have not been included. Moreover, multiple control lines are needed in some systems for single-qubit manipulations. Therefore, it is important to continue to increase the capacity of the cryostat.

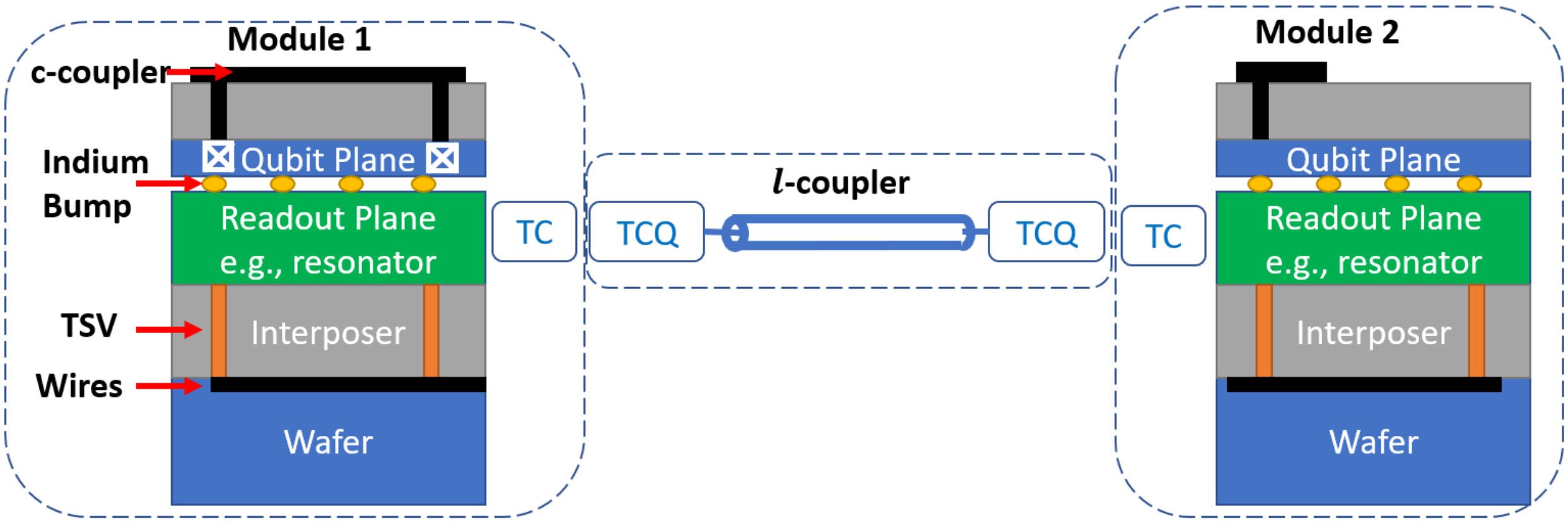

**FIGURE 25.** **A possible large-scale integration scheme based on the talk by Gambetta [210]. Recording is available on the QCE 2025 website and accessed in January, 2026. Note that this figure does not fully reflect the scheme. Qubit and readout microwave circuits are put in different planes and connected through superconducting indium bumps. C-couplers are used to connect non-nearest-neighbor qubits in the same module. TSV and interposer technologies are used for 3D integration. Inter-module quantum connection is accomplished using *l*-coupler. TC stands for tunable coupler. Details of *l*-coupler can be found in [217].**

5. Cryogenic Control Electronics: The number of cables can be reduced, and the throughput can be improved by using cryo-CMOS (i.e., putting controlling CMOS at 4K or below). This has been demonstrated in IBM's Heron device. Using cryo-CMOS AWGs to control the 94 tunable couplers on Heron, the gate fidelity was found to be as high as 99.9% [215].
6. Co-optimization and QPU-HPC Integration: Co-optimization of controlling electronics starting from pulse-level is required [216]. There is also a need for seamless high-performance cluster (HPC) integration with quantum computers for rapid calibration, hybrid applications, and quantum error corrections [217], [218].
7. Design Automation: Seamless EDA tools from *ab initio* to system design are crucial. Emphasis should be put on rapid but accurate simulations. Also, EDA tools need to be able to handle 10k's or even millions of qubits.
8. Architecture: Architecture should be developed to enable high qubit connectivity (e.g., 6 in IBM Loon) and take efficient quantum error correction code implementation into consideration [219].
9. Fabrication uniformity: It is important to improve fabrication uniformity because junction barrier thickness and junction area determine the qubit frequencies. Precise frequency setting is critical to avoid frequency crowding and collision. For example, a frequency variation of less than 0.25% is required [66]. This is a 4X improvement over the current technology.

### *B. Examples*

Here we present three integration examples. In the first example by IBM, it envisages how to expand the existing infrastructure and technology to build a 2000-logical qubit quantum computer through intermodular communications. In the second example, QuantWare emphasizes how to design chiplets and 3D packaging to build high-density qubit modules to facilitate large-scale integration. In the third example, Qolab proposed to use a full-wafer scheme to enable large-scale integration. It is likely that a successful large-scale integration scheme will combine the best from various technologies.

Example 1: In IBM's roadmap, by 2029, fault-tolerant quantum computing with 200 logical qubits and 100 million gates will be available (Starling) [220]. After 2033, a 2000 logical qubit system running with billions of gates is expected to be realized [220]. To realize these, entangling qubits within a chiplet, across chiplets, and even across different dilution refrigerators are necessary [219].

To optimize qubit and microwave circuits independently and to enable 3D integration, a promising approach is to put the qubits and buses on one chip and signal delivery on another chip (breaking the system onto different planes) (Fig. 25). The two chips are bonded together through a superconducting bump (e.g., indium). CMOS technologies such as TSV and silicon interposer are used [221]. The coupling between non-nearest-neighbor qubits within the same chip or across different planes is achieved through c-couplers. c-couplers can couple qubits as far as 14 mm from each other. There is almost no degradation of fidelity when the distance changes from 6 mm ($3\times10^{-3}$) to 14 mm ($5\times10^{-3}$) [215]. To enable chip-to-chip, module-to-module, or even system-to-system communication, $l$-couplers are used [215], [222]. Currently, $l$-couplers operate in a dilution refrigerator as the quantum processor, and long-distance quantum interconnects need to be developed to operate at a higher temperature [223]. A $l$-coupler consists of two tunable coupling qubits (TCQ) and a waveguide (Fig. 25). It has been demonstrated that an $l$-coupler can enable a CNOT between two qubits separated by 0.6m with a fidelity as high as 93.3% [222]. For inter-fridge quantum communication, a transducer that converts microwave photons to optical photons is likely required [223], [224] and is expected to be achieved within 2030 [225]. Eventually, to build an effective architecture, Quantum Network Units (QNUs), which can link quantum processing units and let them share information, would be essential in a large-scale quantum computer [215], [226].

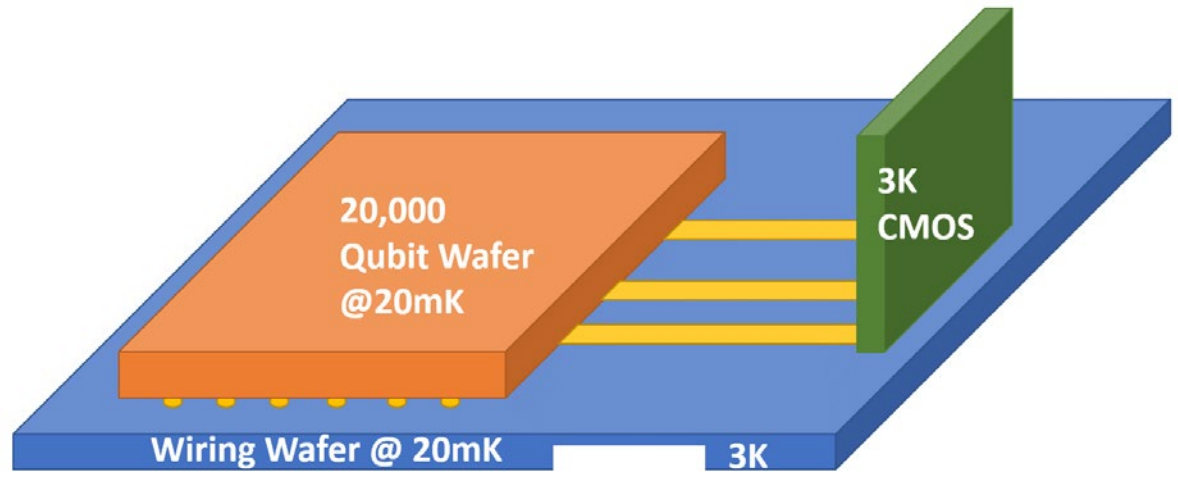

**FIGURE 26. Integration scheme by bonding a full wafer with 20,000 qubits to a wiring wafer at 20mK. It is connected to cryogenic CMOS at 3K. The wiring wafer is selectively thinned to provide thermal insulation between 20mK and 3K. Modified based on [199].**

Finally, to achieve the aforementioned goal for Starling, it is expected that a total of 1,800 $ft^2$ of floor space (fridge and control electronics) is required with a total power of 2 MegaWatt [215].

Example 2: Another modular approach is by QuantWare. It is expected that, by 2028, it will deliver a technology that supports 40,000 input-output lines with 10,000 qubits [227]. In this technology, 3D signal delivery is used to allow a higher packing density of qubits on the qubit plane with less crosstalk. Seamless inter-chip interconnect is used to build a large-scale quantum chip from chiplets.

Example 3: It is envisaged that in 2033, quantum computers may scale to as many as five million qubits [175]. In order to achieve this milestone, in [202], a methodology to build a million-qubit quantum computer is proposed (Fig. 26). Due to coherence loss between dies, this paper suggests fabricating 20,000 qubits on a 300-mm wafer without dicing. It will then be indium bump-bonded to a wiring wafer, which is connected to 3K cryo-CMOS on the other end. The wafer is thinned between the qubit wafer and the cryo-CMOS to increase the thermal resistance so can maintain the temperature gradient across it. Nb is used for wiring, and about 92k wires can be routed. Frequency-multiplexing is used to reduce the number of wires needed. As mentioned earlier, the readout circuit consumes a lot of area and volume due to the circulators and the parametric amplifier. This can be solved using Josephson photomultipliers proposed in [208]. About 5 to 10 such 20k modules can then be tiled together in one dilution refrigerator through couplers. Optical-microwave transducers can then be used to communicate between cryostats if a larger system is needed.

## XI. CONCLUSION

In this paper, we reviewed the history and current status of superconducting qubit quantum computers. It is structured around the five DiVincenzo's criteria. After reviewing the basics of quantum computers, superconductors, and Josephson junctions, we discussed the critical topics related to the criteria. These include qubit designs, two-qubit gates, readout circuits, quantum-limited amplifiers, Purcell filters, and two-level system defects. But a quantum computer is only useful with hundreds of thousands or more qubits. We therefore discussed its possible integration schemes and emphasized the importance of using electronic design automation. Superconducting qubit quantum computers are still in their nascent stage. It is still difficult to predict which qubit technology and integration methodology will win in the end. However, it is almost certain that every endeavor in the research and commercialization of superconducting qubit quantum computers will eventually contribute to its success.

The topics we have discussed are profound and could not be fully covered in one paper. Therefore, the goal of this paper is to serve as a starting point for researchers who want to have a quick and broad, but might be incomplete, overview of the technology.

We do not know if a quantum computer will be useful one day. While investors and companies are pursuing profits, for us, those who love quantum mechanics and those who love knowledge, this is a great opportunity to see the mysterious quantum theory put into practical use. If a quantum computer is ultimately proven not to be useful, those of us who study it for its own sake will not be disappointed and will still carry the memory of a wonderful journey.

## ACKNOWLEDGMENT

This paper is the result of the work supported by the National Science Foundation under Grant No. 2125906, 2024 SJSU RSCA Seed Grant, 2024 SJSU RSCA Equipment Grant, and AMDT Endowment Funding from the College of Engineering, San Jose State University.

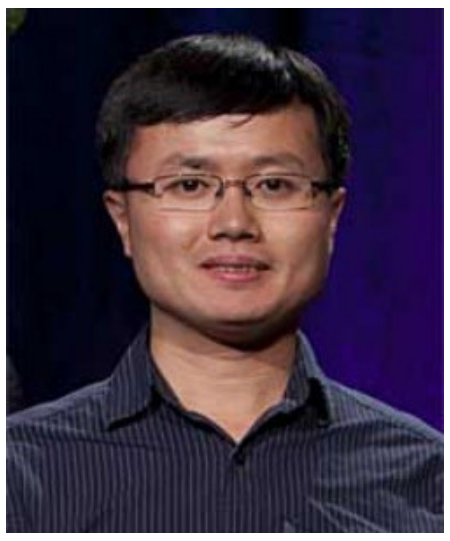

**HIU YUNG WONG** (M'12–SM'17) is a Professor and AMDT Endowed Chair at San Jose State University. He received his Ph.D. degree in Electrical Engineering and Computer Science from the University of California, Berkeley in 2006. From 2006 to 2009, he was an MTS working on technology integration at Spansion. From 2009 to 2018, he was a TCAD Senior Staff Application Engineer at Synopsys.

He received the Industry Sponsored Research Award and ERFA RSCA Award in 2024, the Curtis W. McGraw Research Award from ASEE Engineering Research Council in 2022, the NSF CAREER award and the Newnan Brothers Award for Faculty Excellence in 2021, and the Synopsys Excellence Award in 2010. He is the author of two books, "Introduction to Quantum Computing: From a Layperson to a Programmer in 30 Steps" and "Quantum Computing Architecture and Hardware for Engineers: Step by Step". He is one of the founding faculty members of the Master of Science in Quantum Technology at San Jose State University.

He serves as the editor of the IEEE Electron Device Letters and the IEEE Journal of Electron Devices Society. He is the organizer and technical program committee member in various conferences, such as SISPAD.

His research interests include the application of machine learning in simulation and manufacturing, quantum computing, cryogenic electronics, and wide-bandgap device simulations. His works have produced 2 books, 1 book chapter, more than 130 papers, and 11 patents.